%% file: paper.tex
\documentclass[preprint]{elsarticle}

\pdfoutput=1
 
\usepackage{amsthm}
\usepackage{afterpage}
\usepackage{dsfont}
\usepackage{color}
\usepackage{amssymb}
\usepackage{algorithmic}
\usepackage[ruled,linesnumbered,algo2e]{algorithm2e}
\usepackage{array}
\usepackage{booktabs}
\usepackage{color}
\usepackage{colortbl}
\usepackage{tabularx}
\usepackage{epsfig}
\usepackage{fancybox}
\usepackage{graphicx}
\usepackage{helvet}
\usepackage{listings}
\usepackage{longtable}
\usepackage{multirow}
\usepackage{pifont}
\usepackage{rotate}
\usepackage{amssymb,amsmath}
\usepackage{url}
\usepackage{listings}
\usepackage{paralist}
\usepackage{multirow}
\usepackage{morefloats}
\usepackage[font=footnotesize]{subcaption}
\usepackage{todonotes}
\usepackage{soul}
\usepackage{etex}
\usepackage{graphics}

\newcommand{\MPDeclare}{MP-Declare}

\newcommand{\templateDefinition}[8]{
\begin{small}
	\RestyleAlgo{plain}
	\begin{longtable}{l}
		\toprule
		\textbf{#1} \\
		\midrule
		\endfirsthead
		\toprule
		\textbf{#1} \emph{(continued from previous page)} \\
		\midrule
		\endhead
		\bottomrule \\
		\caption{#2}
		\label{#3}
		\endlastfoot
		\begin{tabular}[t]{@{}l@{}}
			$\textit{template}.\textit{opening}()$ \\
			\begin{algorithm2e}[H] \DontPrintSemicolon
				#4
			\end{algorithm2e}
		\end{tabular} \\
		\midrule
		\begin{tabular}[t]{@{}l@{}}
			$\textit{template}.\textit{fulfillment}(e, \textit{trace}, \textit{pending}, \textit{fulfillments}, T, \varphi_a, \varphi_c, \varphi_{\tau})$ \\
			\begin{algorithm2e}[H] \DontPrintSemicolon
				#6
			\end{algorithm2e}
		\end{tabular} \\
		\midrule
		\begin{tabular}[t]{@{}l@{}}
			$\textit{template}.\textit{violation}(e, \textit{trace}, \textit{pending}, \textit{violations}, T, \varphi_c, \varphi_{\tau})$ \\
			\begin{algorithm2e}[H] \DontPrintSemicolon
				#7
			\end{algorithm2e}
		\end{tabular} \\
		\midrule
		\begin{tabular}[t]{@{}l@{}}
			$\textit{template}.\textit{activation}(e, A, \textit{pending}, \varphi_a)$ \\
			\begin{algorithm2e}[H] \DontPrintSemicolon
				#5
			\end{algorithm2e}
		\end{tabular} \\
		\midrule
		\begin{tabular}[t]{@{}l@{}}
			$\textit{template}.\textit{closing}(\textit{pending}, \textit{fulfillments}, \textit{violations})$ \\
			\begin{algorithm2e}[H] \DontPrintSemicolon
				#8
			\end{algorithm2e}
		\end{tabular} \\
	\end{longtable}
\end{small}
}

\makeatletter
\algocf@newcommand{KwInXX}[1]{%
  \sbox\algocf@inputbox{\hbox{\KwSty{Input}\algocf@typo: }}%
  \ifthenelse{\boolean{algocf@inoutnumbered}}{\relax}{\everypar={\relax}}%
  {\let\\\algocf@newinput\hspace{\wd\algocf@inputbox}\hangindent=\wd\algocf@inputbox\hangafter=\wd\algocf@inputbox#1\par}%
  \algocf@linesnumbered
}
\makeatother

\newcommand{\verify}{\ensuremath{\textit{verify}}}

\soulregister\cite7
\soulregister\ref7

\newtheorem{definition}{Definition}

\title{Conformance Checking Based on Multi-Perspective Declarative Process Models}

\author[unipd]{A.~Burattin}
\ead{burattin@math.unipd.it}

\author[ut]{F.~M.~Maggi}
\ead{f.m.maggi@ut.ee}

\author[unipd]{A.~Sperduti}
\ead{sperduti@math.unipd.it}

\address[unipd]{University of Padua, Italy}
\address[ut]{University of Tartu, Estonia}

\begin{document}

\begin{abstract}
	\input{sections/abstract}
\end{abstract}


\input{sections/macros}

\maketitle
\input{sections/introduction}

\input{sections/related}
\input{sections/preliminaries}
\input{sections/semantics}

\input{sections/algorithms}

\input{sections/benchmarks}
\input{sections/casestudy}
\input{sections/conclusion}

\bibliographystyle{elsarticle-num}
\bibliography{libraryandrea,MarcoBibliography,ale}

\end{document}

%% file: sections/abstract.tex
Process mining is a family of techniques that aim at analyzing business process execution data recorded in event logs. Conformance checking is a branch of this discipline embracing approaches for verifying whether the behavior of a process, as recorded in a log, is in line with some expected behaviors provided in the form of a process model. The majority of these approaches require the input process model to be procedural (e.g., a Petri net). However, in turbulent environments, characterized by high variability, the process behavior is less stable and predictable. In these environments, procedural process models are less suitable to describe a business process. Declarative specifications, working in an open world assumption, allow the modeler to express several possible execution paths as a compact set of constraints. Any process execution that does not contradict these constraints is allowed. One of the open challenges in the context of conformance checking with declarative models is the capability of supporting multi-perspective specifications. In this paper, we close this gap by providing a framework for conformance checking based on \MPDeclare{}, a multi-perspective version of the declarative process modeling language Declare. The approach has been implemented in the process mining tool ProM and has been experimented in three real life case studies.  

%% file: sections/macros.tex
%


\begingroup
\catcode`\~=11
\gdef\urltilde{\lower 0.6ex\hbox{~}}
\endgroup


\newcommand{\A}{\mathcal{A}} \newcommand{\B}{\mathcal{B}}
\newcommand{\C}{\mathcal{C}} \newcommand{\D}{\mathcal{D}}
\newcommand{\E}{\mathcal{E}} \newcommand{\F}{\mathcal{F}}
\newcommand{\G}{\mathcal{G}} \renewcommand{\H}{\mathcal{H}}
\newcommand{\I}{\mathcal{I}} \newcommand{\J}{\mathcal{J}}
\newcommand{\K}{\mathcal{K}} \renewcommand{\L}{\mathcal{L}}
\newcommand{\M}{\mathcal{M}} \newcommand{\N}{\mathcal{N}}
\renewcommand{\O}{\mathcal{O}} \renewcommand{\P}{\mathcal{P}}
\newcommand{\Q}{\mathcal{Q}} \newcommand{\R}{\mathcal{R}}
\renewcommand{\S}{\mathcal{S}} \newcommand{\T}{\mathcal{T}}
\newcommand{\U}{\mathcal{U}} \newcommand{\V}{\mathcal{V}}
\newcommand{\W}{\mathcal{W}} \newcommand{\X}{\mathcal{X}}
\newcommand{\Y}{\mathcal{Y}} \newcommand{\Z}{\mathcal{Z}}
\newcommand{\BE}{{\mathcal{B}, \mathcal{E}}}
\newcommand{\DB}{{\mathcal{DB}}}
\newcommand{\DS}{{\mathcal{DS}}}


\newcommand{\ra}{\rightarrow}
\newcommand{\Ra}{\Rightarrow}
\newcommand{\la}{\leftarrow}
\newcommand{\La}{\Leftarrow}
\newcommand{\lra}{\leftrightarrow}
\newcommand{\Lra}{\Leftrightarrow}
\newcommand{\lora}{\longrightarrow}
\newcommand{\Lora}{\Longrightarrow}
\newcommand{\lola}{\longleftarrow}
\newcommand{\Lola}{\Longleftarrow}
\newcommand{\lolra}{\longleftrightarrow}
\newcommand{\Lolra}{\Longleftrightarrow}
\newcommand{\ua}{\uparrow}
\newcommand{\Ua}{\Uparrow}
\newcommand{\da}{\downarrow}
\newcommand{\Da}{\Downarrow}
\newcommand{\uda}{\updownarrow}
\newcommand{\Uda}{\Updownarrow}

\newcommand{\goto}[1]{\stackrel{#1}{\lora}}


\newcommand{\incl}{\subseteq}
\newcommand{\imp}{\rightarrow}
\newcommand{\dleq}{\dot{\leq}}                   


\newcommand{\per}{\mbox{\bf .}}                  

\newcommand{\cld}{,\ldots,}                      
\newcommand{\ld}[1]{#1 \ldots #1}                 
\newcommand{\cd}[1]{#1 \cdots #1}                 
\newcommand{\lds}[1]{\, #1 \; \ldots \; #1 \,}    
\newcommand{\cds}[1]{\, #1 \; \cdots \; #1 \,}    

\newcommand{\dd}[2]{#1_1,\ldots,#1_{#2}}             
\newcommand{\ddd}[3]{#1_{#2_1},\ldots,#1_{#2_{#3}}}  
\newcommand{\dddd}[3]{#1_{11}\cld #1_{1#3_{1}}\cld #1_{#21}\cld #1_{#2#3_{#2}}}

\newcommand{\ldop}[3]{#1_1 \ld{#3} #1_{#2}}   
\newcommand{\cdop}[3]{#1_1 \cd{#3} #1_{#2}}   
\newcommand{\ldsop}[3]{#1_1 \lds{#3} #1_{#2}} 
\newcommand{\cdsop}[3]{#1_1 \cds{#3} #1_{#2}} 


\newcommand{\quotes}[1]{{\lq\lq #1\rq\rq}}
\newcommand{\set}[1]{\{#1\}}                      
\newcommand{\Set}[1]{\left\{#1\right\}}
\newcommand{\bigmid}{\Big|}
\newcommand{\card}[1]{|{#1}|}                     
\newcommand{\Card}[1]{\left| #1\right|}
\newcommand{\cards}[1]{\sharp #1}
\newcommand{\sub}[1]{[#1]}
\newcommand{\tup}[1]{\langle #1\rangle}            
\newcommand{\Tup}[1]{\left\langle #1\right\rangle}


\newcommand{\inc}[2]{#1\colon #2}



\newcommand{\inter}[1][\I]{(\dom[#1],\Int[#1]{\cdot})}   

\newcommand{\dom}[1][\I]{\Delta^{#1}}  
\newcommand{\Int}[2][\I]{#2^{#1}}      
\newcommand{\INT}[2][\I]{(#2)^{#1}}    


\newcommand{\AND}{\sqcap}
\newcommand{\OR}{\sqcup}
 \newcommand{\NOT}{\neg}
\newcommand{\ALL}[2]{\forall #1 \per #2}
 \newcommand{\SOME}[2]{\exists #1 \per #2}
 \newcommand{\SOMET}[1]{\exists #1}
\newcommand{\ALLRC}{\ALL{R}{C}}
\newcommand{\SOMERC}{\SOME{R}{C}}
\newcommand{\SOMERT}{\SOMET{R}}
\newcommand{\ATLEAST}[2]{(\geq #1 \, #2)}
\newcommand{\ATMOST}[2]{(\leq #1 \, #2)}
\newcommand{\EXACTLY}[2]{(= #1 \, #2)}
 \newcommand{\INV}[1]{#1^{-}}


\newcommand{\QATMOST}[3]{(\leq #1\; \DIAM{#2}{#3})}
\newcommand{\Ident}[1]{(#1)?}


\newcommand{\ISA}{\sqsubseteq}
\newcommand{\EQU}{\equiv}

\newcommand{\QATLEAST}[3]{(\geq #1\, #2 \per #3)}

\newcommand{\ALCQreg}{\mathcal{ALCQ}_\mathit{reg}}
\newcommand{\PDLgm}{\mathit{PDL}_\mathit{gm}}
\newcommand{\ALC}{\mathcal{ALC}}

\newcommand{\Nat}{{\rm I\kern-.23em N}}

\newcommand{\FaCT}{\textsc{FaCT}\xspace}
\newcommand{\Racer}{\textsc{Racer}\xspace}
\newcommand{\SHIQ}{\mathcal{SHIQ}\xspace}

\newcommand{\served}{\mathsf{served}}
\newcommand{\initiated}{\mathsf{initiated}}
\newcommand{\trans}{\mathit{trans}}

\newcommand{\ecard}{\emph{e}-card\xspace}



\newcommand{\eService}{\emph{e-}Service\xspace}
\newcommand{\eServices}{\emph{e-}Services\xspace}
\newcommand{\eservice}{\emph{e-}Service\xspace}
\newcommand{\eservices}{\emph{e-}Services\xspace}
\newcommand{\eGov}{\emph{e-}Government\xspace}
\newcommand{\e}{\emph{e-}}

\newcommand{\Init}{\mathit{Init}}
\newcommand{\KB}{\mathit{KB}}
\newcommand{\VA}{\mathit{VA}}
\newcommand{\VAb}{\mathbf{VA}}

\newcommand{\exec}{\mathit{exec}}
\newcommand{\change}{\mathit{Change}}
\newcommand{\eS}{\mathit{e}\S}
\newcommand{\TeS}{\T^{\eS}}
\newcommand{\Final}{\mathit{Final}}
\newcommand{\Step}{\mathit{Step}}
\newcommand{\Poss}{\mathit{Poss}}
\newcommand{\undef}{\mathit{undef}}
\newcommand{\fin}{\mathit{fin}}
\newcommand{\comp}{\mathit{comp}}

\newcommand{\ttrue}{\mathit{true}}
\newcommand{\ffalse}{\mathit{false}}

\newcommand{\limp}{\rightarrow}
\newcommand{\lequiv}{\leftrightarrow}
\newcommand{\conc}{\mathord{\cdot}}
\newcommand{\eword}{\varepsilon}
\newcommand{\Diam}[2]{\langle #1 \rangle #2}
\newcommand{\Boxx}[2]{[#1]#2}

\newcommand{\myi}{\emph{(i)}\xspace}
\newcommand{\myii}{\emph{(ii)}\xspace}
\newcommand{\myiii}{\emph{(iii)}\xspace}
\newcommand{\myiv}{\emph{(iv)}\xspace}
\newcommand{\myv}{\emph{(v)}\xspace}
\newcommand{\myvi}{\emph{(vi)}\xspace}

\newcommand{\moved}{\mathit{moved}}
\newcommand{\itext}{{\mathit{ext}}}
\newcommand{\itint}{{\mathit{int}}}

\newcommand{\ES}[1]{{#1}^\itext}
\newcommand{\IS}[1]{{#1}^\itint}
\newcommand{\EA}[1]{A^\itext_{#1}}
\newcommand{\IA}[1]{A^\itint_{#1}}
\newcommand{\CG}{CG}
\newcommand{\NCG}{NCG}
\newcommand{\MNCG}{MNCG}
\newcommand{\CHOOSE}{choose(\cdot)}
\newcommand{\NDSP}{NDSP}
\newcommand{\NDS}{NDS}
\newcommand{\MNDS}{MNDS}
\newcommand{\PLAN}{\mbox{\textit{SP}}}

\newcommand{\figureDraft}{
\begin{tabular}{|p{7cm}|}
\hline
\vspace{4cm}\\
\hline
\end{tabular}
}

\newcommand{\ttt}{{\mathit{tt}}}
\newcommand{\fff}{{\mathit{ff}}}
\newcommand{\uuu}{{\mathit{uu}}}
\newcommand{\transA}[3]{\{#1\}#2\{#3\}}
\newcommand{\transT}[2]{\{#1\}#2}

\newcommand{\reminder}[1]{\marginpar{\mbox{$<NEW>$}} [[{\it\small #1}]]}

\newcommand{\bsy}{\boldsymbol}
\newcommand{\certans}{\rhd}

\newcommand{\Goto}[1]{\stackrel{#1}{\Longrightarrow}}
\newcommand{\mf}{\mathfrak}
\newcommand{\msf}{\mathsf}
\newcommand{\homeq}{\stackrel{h}{=}}
\newcommand{\eqmnr}{\stackrel{mnr}{=}}
\newcommand{\sseqmnr}{\stackrel{mnr}{\subseteq}}
\newcommand{\ssmnr}{\stackrel{mnr}{\subset}}


\newcommand{\Moda}[1]{#1_{\V}^\mf{A}}     
\newcommand{\Modax}[2]{#1_{\V #2}^\mf{A}} 
\newcommand{\MODA}[1]{(#1)_{\V}^\mf{A}}   
\newcommand{\MODAX}[2]{(#1)_{\V #2}^\mf{A}} 

\newcommand{\Modat}[2]{#1_{\V}^{#2}}     
\newcommand{\MODAT}[2]{(#1)_{\V}^{#2}}   

\renewcommand{\varsigma}{\mathit{R}}

\newcommand{\Amax}{\A_{\mathit{max}}}

\newcommand{\ex}[1]{\mathsf{#1}}
\newcommand{\actex}[1]{\mathsf{#1}}

\newcommand{\const}[1]{\C_{#1}}

\newcommand{\brho}{\bsy{\rho}}
\newcommand{\bpi}{\bsy{\pi}}

\newcommand{\ans}[2][]{\mathit{ans}_{#1}(#2)}
\newcommand{\rew}[1]{\mathit{rew}(#1)}
\newcommand{\DO}[1]{\mathit{do}(#1)}
\newcommand{\GOTO}[1][\mf{A}]{\mathrel{\Ra_{#1}}}

\newcommand{\conj}{\mathit{conj}}

\newcommand{\homom}[4][\C]{#3 \mathrel{\ra^{#1}_{#2}} #4}
\newcommand{\map}[2]{#1 \rightsquigarrow #2}
\newcommand{\carule}[2]{#1 \mapsto #2}

\newcommand{\dllite}{\textit{DL-Lite}\xspace}
\newcommand{\dllitef}{\textit{DL-Lite}\ensuremath{_{\mathcal{F}}}\xspace}
\newcommand{\dlliter}{\textit{DL-Lite}\ensuremath{_{\mathcal{R}}}\xspace}
\newcommand{\dllitea}{\textit{DL-Lite}\ensuremath{_{\mathcal{A}}}\xspace}


\newcommand{\LTLthree}{{\sc ltl}$_3$\xspace}
\newcommand{\bs}{\boldsymbol}
\newcommand{\univproj}[1]{#1_{\downarrow \forall}}
\newcommand{\existproj}[1]{#1_{\downarrow \exists}}
\newcommand{\bsl}{\backslash}
\newcommand{\prop}{\P rop}
\newcommand{\pfunct}{\mathsf{p}}
\newcommand{\ffunct}{\mathsf{f}}
\newcommand{\rfunct}{\mathsf{r}}

\newcommand{\old}{\mathsf{old}}
\newcommand{\upsphi}{\upsilon^{\Phi}}
\newcommand{\upsnotphi}{\upsilon^{\neg \Phi}}
\newcommand{\epsphi}{\epsilon^{\Phi}}
\newcommand{\epsnotphi}{\epsilon^{\neg \Phi}}

\newcommand{\id}{\textsc{id}\xspace}
\newcommand{\ids}{\textsc{id}s\xspace}

\newcommand{\pid}{p-\textsc{id}\xspace}
\newcommand{\pids}{p-\textsc{id}s\xspace}

\newcommand{\uid}{$\forall$-\textsc{id}\xspace}
\newcommand{\uids}{$\forall$-\textsc{id}s\xspace}

\newcommand{\eid}{$\exists$-\textsc{id}\xspace}
\newcommand{\eids}{$\exists$-\textsc{id}s\xspace}

\newcommand{\cid}{c-\textsc{id}\xspace}
\newcommand{\cids}{c-\textsc{id}s\xspace}

\newcommand{\coddid}{Codd-\textsc{id}\xspace}
\newcommand{\coddids}{Codd-\textsc{id}s\xspace}

\newcommand{\offid}{offline-\textsc{id}\xspace}
\newcommand{\offids}{offline-\textsc{id}s\xspace}

\newcommand{\prun}{p-run\xspace}

\newcommand{\urun}{$\forall$-run\xspace}

\newcommand{\erun}{$\exists$-run\xspace}

\newcommand{\crun}{c-run\xspace}
\newcommand{\cruns}{c-runs\xspace}

\newcommand{\coddrun}{Codd-run\xspace}

\newcommand{\offrun}{offline-run\xspace}

\newcommand{\Buchi}{B\"{u}chi\xspace}
\renewcommand{\=}{\! = \!}

\newcommand{\nv}{\star}

\newcommand{\coddify}{\mathsf{coddify}}
\newcommand{\rep}{\mathsf{rep}}
\newcommand{\poss}{\mathsf{poss}}

\newcommand{\cin}{\! \in \!}

\newcommand{\FO}{{\sc fo}\xspace}
\newcommand{\LT}{{\sc lt}$_f$\xspace}
\newcommand{\EC}{{\sc ec}\xspace}
\newcommand{\FOLTL}{{\sc fo-ltl}\xspace}
\newcommand{\RVLTL}{{\sc rv-ltl}\xspace}
\newcommand{\LTLf}{{\sc ltl}$_f$\xspace}
\newcommand{\LTLi}{{\sc ltl}\xspace}
\newcommand{\LDL}{{\sc ldl}\xspace}
\newcommand{\add}{\mathit{ADD}}
\newcommand{\del}{\mathit{DEL}}
\newcommand{\LDLf}{{\sc ldl}$_f$\xspace}
\newcommand{\RE}{{\sc re}$_f$\xspace}
\newcommand{\PDL}{{\sc pdl}\xspace}
\newcommand{\FOLf}{{\sc fol}$_f$\xspace}
\newcommand{\MSOf}{{\sc mso}$_f$\xspace}
\newcommand{\FOL}{{\sc fol}\xspace}
\newcommand{\MSO}{{\sc mso}\xspace}
\newcommand{\AFW}{{\sc afw}\xspace}
\newcommand{\NFA}{{\sc nfa}\xspace}
\newcommand{\DFA}{{\sc dfa}\xspace}
\newcommand{\CTL}{{\sc ctl}\xspace}
\newcommand{\QLTL}{{\sc qltl}\xspace}
\newcommand{\muLTL}{$\mu${\sc ltl}\xspace}
\newcommand{\declare}{{\sc declare}\xspace}
\newcommand{\fol}{\mathit{fol}}
\newcommand{\f}{\mathit{f}}
\newcommand{\g}{\mathit{g}}
\newcommand{\re}{\mathit{re}}

\newcommand{\Next}{\raisebox{-0.27ex}{\LARGE$\circ$}}
\newcommand{\Wnext}{\raisebox{-0.27ex}{\LARGE$\bullet$}}
\newcommand{\Until}{\mathop{\U}}
\newcommand{\Release}{\mathop{\R}}
\newcommand{\Wuntil}{\mathop{\W}}

\newcommand{\temptrue}{\mathit{temp\_true}}
\newcommand{\tempfalse}{\mathit{temp\_false}}
\newcommand{\Last}{\mathit{Last}}
\newcommand{\Ended}{\mathit{Ended}}
\newcommand{\length}{\mathit{length}}
\newcommand{\last}{\mathit{last}}
\newcommand{\nnf}{\mathit{nnf}}
\newcommand{\CL}{\mathit{CL}}
\newcommand{\MU}[2]{\mu #1.#2}
\newcommand{\NU}[2]{\nu #1.#2}
\newcommand{\BOX}[1]{ [#1]}
\newcommand{\DIAM}[1]{\langle #1 \rangle}

%% file: sections/introduction.tex
\section{Introduction}
\label{sec:intro}

The need to develop information systems able to fully support business processes of companies, and organizations in general, is becoming more and more urgent because of the fast pace of change in markets. Such dynamic markets impose frequent modifications and updates to business processes, leading to a constant decrease, in terms of temporal span, to the life-cycle of a business process definition. In this context, one very important functionality that any process-aware information system should be able to support is {\it conformance checking}, i.e., the ability to verify whether the actual flow of work is conformant with the intended business process model. This is especially true in the case of very complex processes, where the adoption of an imperative formalism to represent it, such as Petri Nets \cite{Murata89} or BPM Notation \cite{BPMN20}, may lead to so much intricate workflows (so called ``spaghetti''-like workflows) to become basically impossible to even properly visualize the process for human inspection.

Early works in conformance checking (e.g., \cite{DBLP:journals/tosem/CookW99,DBLP:journals/is/RozinatA08,aalst_book}) mainly focused on the control-flow perspective in the context of imperative models, i.e., on the functional dependencies among performed activities/tasks in the process, while abstracting from time constraints, data dependencies, and resource assignments. These works were mainly based on replaying the log on the model to compute, according to the proposed approach, the fraction of events or traces in the log that can be replayed by the model. An evolution of these approaches is given by align-based approaches, where the conformance checking is performed by aligning both the modeled behavior and the behavior observed in the log (e.g. \cite{DBLP:conf/edoc/AdriansyahDA11}).
Only recently, approaches able to deal with multiple perspectives have been developed \cite{DeLeoni2013,BPM-14-07}, as well as approaches that aim at being computationally efficient via a problem decomposition strategy \cite{DBLP:conf/apn/Aalst12,DBLP:journals/dpd/Aalst13,DBLP:conf/otm/LeoniMCA14,DBLP:journals/is/Munoz-GamaCA14}.

In the case in which the process in consideration is complex, however, it is much better to use a declarative formalism, such as Declare \cite{declareCSRD09,declare_www,maja-declare-edoc07}, to represent a set of constraints that must be satisfied throughout the process execution.
In this way, the ``spaghetti''-like workflows are avoided, and the obtained model is flexible enough to allow all behaviors that do not violate the defined constraints. Conformance checking approaches based on the control-flow perspective have been defined for declarative models as well (e.g. \cite{Chesani2009,Montali2010a,Burattin2012}). More recently the additional data perspective has been considered in \cite{DBLP:conf/sac/MontaliCMM13,Borrego2014}, even if in these works the data perspective is not fully integrated with the control flow perspective.
Efficient and fully integrated multi-perspective conformance checking proposals for declarative models, however, are still missing.

In this paper, we aim at closing this gap by proposing a multi-perspective approach based on Declare where it is possible to define multi-perspective constraints jointly considering data, temporal, and control flow perspectives. In order to allow that,
we formally define Multi-Perspective Declare (\MPDeclare), an augmented version of Declare where, thanks to the use of  Metric First-Order Linear Temporal Logic,  it is possible to define activation, correlation, and time conditions to build constraints over traces.

A nice feature of \MPDeclare\ is that, by construction, it allows the user to efficiently perform conformance checking over event logs. In fact, we show that it is possible to define a conformance checking algorithmic framework operating on constraint templates, that is linear  in  the number of traces, constraints, and in the number of events of each trace. Conformance checking for a specific template is then obtained via definition of  template-dependent procedures within the framework, whose time complexity depends on the actual template. Overall, however, the time complexity is upper bounded in the worst case by a quadratic function.

We assess the validity of the proposed approach both on artificial and real event logs. Controlled artificial data, involving logs containing up to 5 million events, are used to prove the scalability of the proposed approach, while real event logs  generated by three real business processes are used to demonstrate the expressivity and flexibility of constraints defined via \MPDeclare.

%% file: sections/related.tex
\section{Related Work}
\label{sec:related}

The scientific literature reports several works in the field of conformance checking \cite{VanderAalstBook}. Typically, the term \emph{conformance checking} refers to the comparison of observed behaviors -- as recorded in an event log -- with respect to a process model. In the past, most of the conformance checking techniques were based on procedural models. State of the art examples of these approaches are reported in \cite{DeLeoni2013,Adriansyah2014,DBLP:conf/otm/LeoniMCA14,DBLP:journals/is/Munoz-GamaCA14}.

In recent years, an increasing number of researchers are focusing on the conformance checking with respect to declarative models.
For example, in \cite{Chesani2009}, an approach for compliance checking with respect to \emph{reactive business rules} is proposed. Rules, expressed using Condec \cite{Pesic2006}, are mapped to Abductive Logic Programming, and Prolog is used to perform the validation. The approach has been extended in \cite{Montali2010a}, by mapping constraints to LTL, and evaluating them using automata. The entire work has been contextualized into the service choreography scenario.

Runtime monitoring for compliance checking has been studied also based on MFOTL, as reported in \cite{Basin2013,Basin2013a}. In these cases, the focus is on security policy monitoring. On the one side the authors try to enforce security policies, on the other they perform monitoring. In order to enforce security policies, it is necessary to distinguish between \emph{controllable} and \emph{observable} activities and, under specific circumstances, terminate the systems in order to prevent policy violations. Concerning the monitoring, authors identified fragments of the used logic, to describe security policies insensitive with respect to the ordering of actions with equal timestamps. The authors assume to perform monitoring in a distributed systems, which have synchronized clocks with limited precision.

Another application domain that researchers used to assess the applicability of conformance checking techniques is the medical domain. In particular, Grando et al. \cite{Grando2012,Grando2013} used Declare to model medical guidelines and to provide semantic (i.e., ontology-based) conformance checking measures. However, in this analysis neither data nor time perspectives are taken into account.

In \cite{Burattin2012}, the authors report an approach that can be used to evaluate the conformance of a log with respect to a Declare model. In particular, their algorithms compute, for each trace, whether a Declare constraint is violated or fulfilled. Using these statistics the approach allows the user to evaluate the ``healthiness'' of the log. The approach is based on the conversion of Declare constraints into automata and, using a so-called ``activation tree'', it is able to identify violations and fulfillments. The approach described in this work does not take into account the data and time perspective, but only the control-flow is analyzed.

The work described in \cite{Leoni2012,DeLeoni2014} consists in converting a Declare model into an automaton and perform conformance checking of a log with respect to the generated automaton. The conformance checking approach is based on the concept of ``alignment'' and as a result of the analysis each trace is converted into the most similar trace that the model accepts.

In a recent work, reported in \cite{Borrego2014}, the data perspective for conformance checking with Declare is expressed in terms of conditions on global variables disconnected from the specific Declare constraints expressing the control flow. This work does not take the temporal perspective into account. In contrast, we provide a formal semantics in which the data perspective, the temporal perspective and the control flow are connected with each others.

%% file: sections/preliminaries.tex
\newcommand{\lnext}{\ensuremath{\mathbf{X}}}
\newcommand{\lwnext}{\ensuremath{\mathbf{\bar{X}}}}
\newcommand{\luntil}{\ensuremath{\mathbf{U}}}
\newcommand{\lsince}{\ensuremath{\mathbf{S}}}
\newcommand{\lrelease}{\ensuremath{\mathbf{R}}}
\newcommand{\lwuntil}{\ensuremath{\mathbf{W}}}
\newcommand{\lglobally}{\ensuremath{\mathbf{G}}}
\newcommand{\lfuture}{\ensuremath{\mathbf{F}}}
\newcommand{\tnext}{\ensuremath{\mathbf{X}_{I}}}
\newcommand{\twnext}{\ensuremath{\mathbf{\bar{X}_I}}}
\newcommand{\tuntil}{\ensuremath{\mathbf{U}_{I}}}
\newcommand{\tsince}{\ensuremath{\mathbf{S}_{I}}}
\newcommand{\trelease}{\ensuremath{\mathbf{R}_{I}}}
\newcommand{\tglobally}{\ensuremath{\mathbf{G}_{I}}}
\newcommand{\lonce}{\ensuremath{\mathbf{O}}}
\newcommand{\tonce}{\ensuremath{\mathbf{O}_{I}}}
\newcommand{\lyesterday}{\ensuremath{\mathbf{Y}}}
\newcommand{\tyesterday}{\ensuremath{\mathbf{Y}_{I}}}
\newcommand{\lhistorically}{\ensuremath{\mathbf{H}}}
\newcommand{\thistorically}{\ensuremath{\mathbf{H}_{I}}}
\newcommand{\tfuture}{\ensuremath{\mathbf{F}_{I}}}
\newcommand{\true}{\ensuremath{\mbox{true}}}
\newcommand{\false}{\ensuremath{\mbox{false}}}

\section{Preliminaries}
\label{sec:preliminaries}

In this section, we present the fundamental concepts required to understand the rest of the paper.


\subsection{Process Mining and XES}
\label{sec:mining}

The basic idea behind process mining is to discover, monitor and improve processes by extracting knowledge from data that is available in today's systems \cite{aalst_book}. The starting point for process mining is an event log. XES (eXtensible Event Stream) \cite{XES-standard-2013,Verbeek10} has been developed as the standard for storing, exchanging and analyzing event logs.

Each event in a log refers to an activity (i.e., a well-defined step in some process) and is related to a particular case (i.e., a process instance). The events belonging to a case are ordered with respect to their execution times. Hence, a case (i.e., a trace) can be viewed as a sequence of events. Event logs may store additional information about events such as the resource (i.e., person or device) executing or initiating the activity, the timestamp of the event, or data elements recorded with the event. In XES, data elements can be event attributes, i.e., data produced by the activities of a business process and case attributes, namely data that are associated to a whole process instance. In this paper, we assume that all attributes are globally visible and can be accessed/manipulated by all activity instances executed inside the case.

\subsection{Metric First Order Temporal Logic} \label{sec:mtl}
\label{sec:mfotl}

In this paper, we use Metric First Order Temporal Logic (MFOTL) first introduced in \cite{Chomicki:1995:ECT:210197.210200}. MFOTL extends propositional metric temporal
logic \cite{MTL} to merge the expressivity of first-order logic together with the MTL temporal modalities.
We deal with a fragment of MFOTL where all traces are finite.

In the following, we call ``structure'' a triple $D = (\Delta,\sigma,\iota)$. $\Delta$ is the domain of the structure, i.e., an arbitrary set. $\sigma$ is the signature of the structure, i.e., a triple $\sigma = (C,R,a)$, where $C$ is a set of constant symbols, $R$ is a set of relational symbols, and $a$ is a function that specify the arity of each relational symbol.
$\iota$ is the interpretation function of the structure that assigns a meaning to all the symbols in $\sigma$ over the domain $\Delta$.

\begin{definition}[Timed temporal structure]\label{def:structure}
	A \emph{timed temporal structure} over the signature $\sigma = (C,R,a)$ is a pair $(D,\tau)$ where $D$ is a finite sequence of structures $D=(D_1,\dots,D_n)$ and $\tau=(\tau_1,\dots,\tau_n)$ is a finite sequence of timestamps with $\tau_i \in \mathds{N}$.\footnote{Note that every timestamp available in a XES log can be translated into an integer.} $D$ is assumed to have constant domains, i.e., $\Delta_i=\Delta_{i+1}$, for all $1\leq i < n$. Each constant symbol in $C$ has an interpretation that does not vary over the time. The sequence of timestamps $\tau$ is monotonically increasing, i.e., $\tau_i \leq \tau_{i+1}$, for all $1\leq i < n$.
\end{definition}

We indicate with $I=[a,b)$ an interval, where $a \in \mathds{N}$ and $b \in \mathds{N}\cup\{\infty\}$, and with $V$ a set of variables.
To express MFOTL formulas, we use the syntax:
\begin{definition}[MFOTL Syntax]\label{def:tltl}
	Formulas of MFOTL over a signature $\sigma = (C,R,a)$ are given by the grammar
\begin{align*}
\phi::=t_1 \approx t_2\;|\;r(t_1, \dots, t_{a(r)})\;|\;\neg\phi\;|\;\phi_1\land\phi_2\;|\;\exists x.\phi\;|\;\tnext\phi\;
|\;\phi_1\tuntil\phi_2\;|\;\tyesterday\phi\;|\;\phi_1\tsince\phi_2\;
\end{align*}
where $\phi,\phi_1,\phi_2\in$MFOTL, $I=[a,b)$ is an interval, $r$ is an element of $R$, $x$ ranges over $V$, and $t_1, t_2,\dots$ belong to $V \cup C$.
\end{definition}

A valuation is a mapping $v:V \rightarrow \Delta$. With abuse of notation, if $c$ is a constant symbol in $C$, we say that $v(c)=c$.
For a valuation $v$, a variable $x\in V$, and $d\in\Delta$, $v[x/d]$ is the valuation that maps $x$ to $d$ and leaves unaltered the valuation of the other variables.

\begin{definition}[MFOTL Semantics]\label{def:tltl}
Given $(D,\tau)$ a timed temporal structure over the signature $\sigma = (C,R,a)$ with $D=(D_1,\dots,D_n)$, $\tau=(\tau_1,\dots,\tau_n)$, $\phi$ a formula over $S$, v a valuation, and $1\leq i\leq n$, we define $(D, \tau, v, i) \vDash \phi$ as follows:
$$
\begin{array}{rcl}
	(D, \tau, v, i) \vDash t \approx t' & \text{iff} & v(t) = v(t') \\
	(D, \tau, v, i) \vDash r(t_1, \dots, t_{a(r)}) & \text{iff} & (v(t_1), \dots, v(t_{a(r)}))) \in \iota(r) \\
	(D, \tau, v, i) \vDash (\neg \phi_1) & \text{iff} & (D, \tau, v, i) \nvDash \phi_1 \\
	(D, \tau, v, i) \vDash \phi_1 \wedge \phi_2 & \text{iff} & (D, \tau, v, i) \vDash \phi_1 \text{ and } (D, \tau, v, i) \vDash \phi_2 \\
	(D, \tau, v, i) \vDash \exists x.\phi_1 & \text{iff} & (D, \tau, v[x/d], i) \vDash \phi_1 \text{, for some } d \in \Delta \\
	(D, \tau, v, i) \vDash \tyesterday \phi_1 & \text{iff} & i > 1, \tau_i - \tau_{i-1} \in I \text{, and } (D, \tau, v, i-1) \vDash \phi_1 \\
	(D, \tau, v, i) \vDash \tnext \phi_1 & \text{iff} & i < n, \tau_{i+1} - \tau_i \in I \text{ and } (D, \tau, v, i+1) \vDash \phi_1 \\
	(D, \tau, v, i) \vDash \phi_1 \tsince \phi_2 & \text{iff} & \text{for some } j \leq i, \tau_i - \tau_j \in I, \\
		& & (D, \tau, v, j) \vDash \phi_2 \text{ and } (D, \tau, v, k) \vDash \phi_1 \\
		& & \text{for all } k \in [j+1, i+1) \\
	(D, \tau, v, i) \vDash \phi_1 \tuntil \phi_2 & \text{iff} & \text{for some } j \geq i, \tau_j - \tau_i \in I, \\
		& & (D, \tau, v, j) \vDash \phi_2 \text{ and } (D, \tau, v, k) \vDash \phi_1 \\
		& & \text{for all } k \in [j, i) \\
\end{array}
$$
\end{definition}

We add syntactic sugar for the normal connectives, such as
$true \equiv \exists x.x \approx x$, $\phi_1\lor\phi_2 \equiv \neg(\neg\phi_1\land \neg\phi_2)$, $\forall x. \phi \equiv \neg \exists x.\neg\phi$ $\phi_1\rightarrow\phi_2\equiv(\neg\phi_1)\lor\phi_2$ and
$\phi_1\leftrightarrow\phi_2\equiv(\phi_1\rightarrow\phi_2)\land(\phi_2\rightarrow\phi_1)$.
We also add temporal syntactic sugar,
$\tfuture\psi\equiv\true\tuntil\psi$ (timed future operator),
$\tglobally\psi\equiv\neg(\tfuture(\neg\psi))$ (timed globally operator), $\tonce\psi\equiv\true\tsince\psi$ (timed once operator) and
$\thistorically\psi\equiv\neg(\tonce(\neg\psi))$ (timed historically
operator).
The non-metric variants of the temporal operators are obtained by specifying $I=[0,\infty)$.

\begin{table}[tb]
\caption{Semantics for some Declare templates. \label{tbl:timed-ltl}}
\centering
\begin{tabular}{llc}
\toprule
\textbf{Template}    & \textbf{LTL semantics} &  \textbf{Activation}\\
\midrule
responded existence  & $\lglobally(A \rightarrow (\lonce B \vee \lfuture B))$ & $A$\\
\midrule
response &  $\lglobally(A \rightarrow \lfuture B)$ & $A$ \\
alternate response  & $ \lglobally(A \rightarrow \lnext(\neg A \luntil B))$ & $A$\\
chain response &  $\lglobally(A \rightarrow \lnext B)$& $A$ \\
\midrule
precedence &  $\lglobally(B \rightarrow \lonce A)$ & $B$\\
alternate precedence & $\lglobally(B \rightarrow \lyesterday(\neg B \lsince A ))$ & $B$\\
chain precedence & $\lglobally(B  \rightarrow \lyesterday A)$ & $B$\\
\midrule
not responded existence  &
$\lglobally(A \rightarrow \neg (\lonce B \vee \lfuture B ))$ & $A$\\
not response  & $\lglobally(A \rightarrow \neg \lfuture B )$ & $A$\\
not precedence & $\lglobally(B \rightarrow \neg \lonce A )$ & $B$\\
not chain response  & $\lglobally(A \rightarrow \neg \lnext B )$ & $A$\\
not chain precedence  & $\lglobally(B \rightarrow \neg \lyesterday A )$ & $B$\\
\bottomrule
\end{tabular}
\end{table}

\subsection{Declare}
\label{sec:declare}
Declare is a declarative process modeling language originally introduced by
Pesic and van der Aalst in \cite{declareCSRD09,declare_www,maja-declare-edoc07}. Instead of explicitly
specifying the flow of the interactions among process activities, Declare
describes a set of constraints that must be satisfied throughout the process
execution. The possible orderings of activities are implicitly specified by
constraints and anything that does not violate them is possible during execution. In
comparison with procedural approaches that produce ``closed'' models, i.e., all
that is not explicitly specified is forbidden, Declare models are ``open'' and
tend to offer more possibilities for the execution. In this way, Declare enjoys
flexibility and is very suitable
 for highly dynamic processes characterized by high complexity and variability
 due to the turbulence and the changeability of their execution environments.

A Declare model consists of a set of constraints applied to
activities. Constraints, in turn, are based on templates. Templates
are patterns that define parameterized classes of properties, and
constraints are their concrete instantiations  (we indicate template parameters with capital letters and concrete activities in their instantiations with lower case letters).
They have a graphical representation understandable to the user and
 their semantics can be formalized using different logics
 \cite{Montali2010:Choreographies}, the main one being LTL over finite
 traces, making them verifiable and executable.  Each constraint
 inherits the graphical representation and semantics from its
template.
\tablename~\ref{tbl:timed-ltl} summarizes some Declare templates (the
reader can refer to \cite{declareCSRD09} for a full description of the
language).

The \emph{responded existence} template specifies that if \emph{A} occurs, then
\emph{B} should also occur (either before or after \emph{A}). The
\emph{response} template specifies that when \emph{A} occurs, then \emph{B}
should eventually occur after \emph{A}. The \emph{precedence} template indicates
that \emph{B} should occur only if \emph{A} has occurred before.
Templates \emph{alternate response} and \emph{alternate precedence} strengthen
the response and precedence templates respectively by specifying that activities
must alternate without repetitions in between. Even stronger ordering relations
are specified by templates \emph{chain response} and \emph{chain precedence}.
These templates require that the occurrences of $A$ and $B$
are next to each other.
Declare also includes some negative constraints to explicitly forbid the execution of
activities. The \emph{not responded existence} template indicates that if
\emph{A} occurs in a process instance, \emph{B} cannot occur in the same
instance. According to the \emph{not response} template any occurrence of
\emph{A} cannot be eventually followed by \emph{B}, whereas the \emph{not
precedence} template requires that any occurrence of \emph{B} is not preceded by
\emph{A}. Finally, according to the \emph{not chain response} and \emph{not
chain precedence}, \emph{A} and \emph{B} cannot occur one immediately after the
other.

The major benefit of using templates is that analysts do not have to be aware of the underlying logic-based formalization to understand the models. They work with the graphical representation of templates, while the underlying formulas remain hidden.
Declare is very suitable for specifying compliance models that are used to check if the behavior of a
system complies with desired regulations. The compliance model defines the constraints
related to a single process instance, and the overall expectation is that all
instances comply with the model. Consider, for example, the \emph{response} constraint $\lglobally(a \rightarrow \lfuture b)$. This constraint indicates that if $a$ {\it occurs}, $b$ must
eventually {\it follow}.
Therefore, this constraint is satisfied for traces such as $\textbf{t}_1$ =
$\langle a, a, b, c \rangle$, $\textbf{t}_2 = \langle b,
b, c, d \rangle$, and $\textbf{t}_3 = \langle a, b,
c, b \rangle$, but not for $\textbf{t}_4 = \langle a, b,
a, c \rangle$ because, in this case, the second instance of $a$ is not followed by a $b$. Note that, in $\textbf{t}_2$,
the considered response constraint is satisfied in a trivial way because $a$ never occurs.
In this case, we say that the constraint is \emph{vacuously
satisfied}~\cite{kupf:vacu03}. In \cite{Burattin2012}, the authors introduce
the notion of \emph{behavioral vacuity detection} according to which a
constraint is non-vacuously satisfied in a trace when it is activated in that
trace. An \emph{activation} of a constraint in a trace is an event whose
occurrence imposes, because of that constraint, some obligations on other events (targets)
in the same trace. For example, $a$ is an activation for the \emph{response}
constraint $\lglobally(a \rightarrow \lfuture b)$ and $b$ is a target, because the execution of $a$ forces $b$ to be executed, eventually. In \tablename~\ref{tbl:timed-ltl}, for each template the corresponding activation is specified.

An activation of a constraint can be a \emph{fulfillment} or a \emph{violation}
for that constraint. When a trace is perfectly compliant with respect to a constraint,
every activation of the constraint in the trace leads to a fulfillment.
Consider, again, the response constraint $\lglobally(a \rightarrow \lfuture b)$. In trace $\textbf{t}_1$, the
constraint is activated and fulfilled twice, whereas, in trace $\textbf{t}_3$,
the same constraint is activated and fulfilled only once. On the other hand,
when a trace is not compliant with respect to a constraint, an activation of the
constraint in the trace can lead to a fulfillment but also to a violation (at
least one activation leads to a violation). In trace $\textbf{t}_4$, for
example, the response constraint $\lglobally(a \rightarrow \lfuture b)$ is activated twice, but the first activation leads to a
fulfillment (eventually $b$ occurs) and the second activation
leads to a violation ($b$ does not occur subsequently). An algorithm to
discriminate between fulfillments and violations for a constraint in a trace is
presented in \cite{Burattin2012}. \tablename~\ref{tbl:timed-ltl} reports the activations for the main Declare templates.

In \cite{Burattin2012}, the authors define two metrics to measure the conformance of an event log with respect to~a constraint in terms of violations and fulfillments, called \emph{violation ratio} and \emph{fulfillment ratio} of the constraint in the log. These metrics are valued 0 if the log contains no activations of the considered constraint. Otherwise, they are evaluated as the percentage of violations and fulfillments of the constraint over the total number of activations.

Tools implementing process mining approaches based on Declare are presented in \cite{DBLP:conf/bpm/Maggi13}. The tools are implemented as plug-ins of the process mining framework ProM. 

%% file: sections/semantics.tex
\pdfoutput=1

\begin{figure}[t!]
	\includegraphics[width=\textwidth]{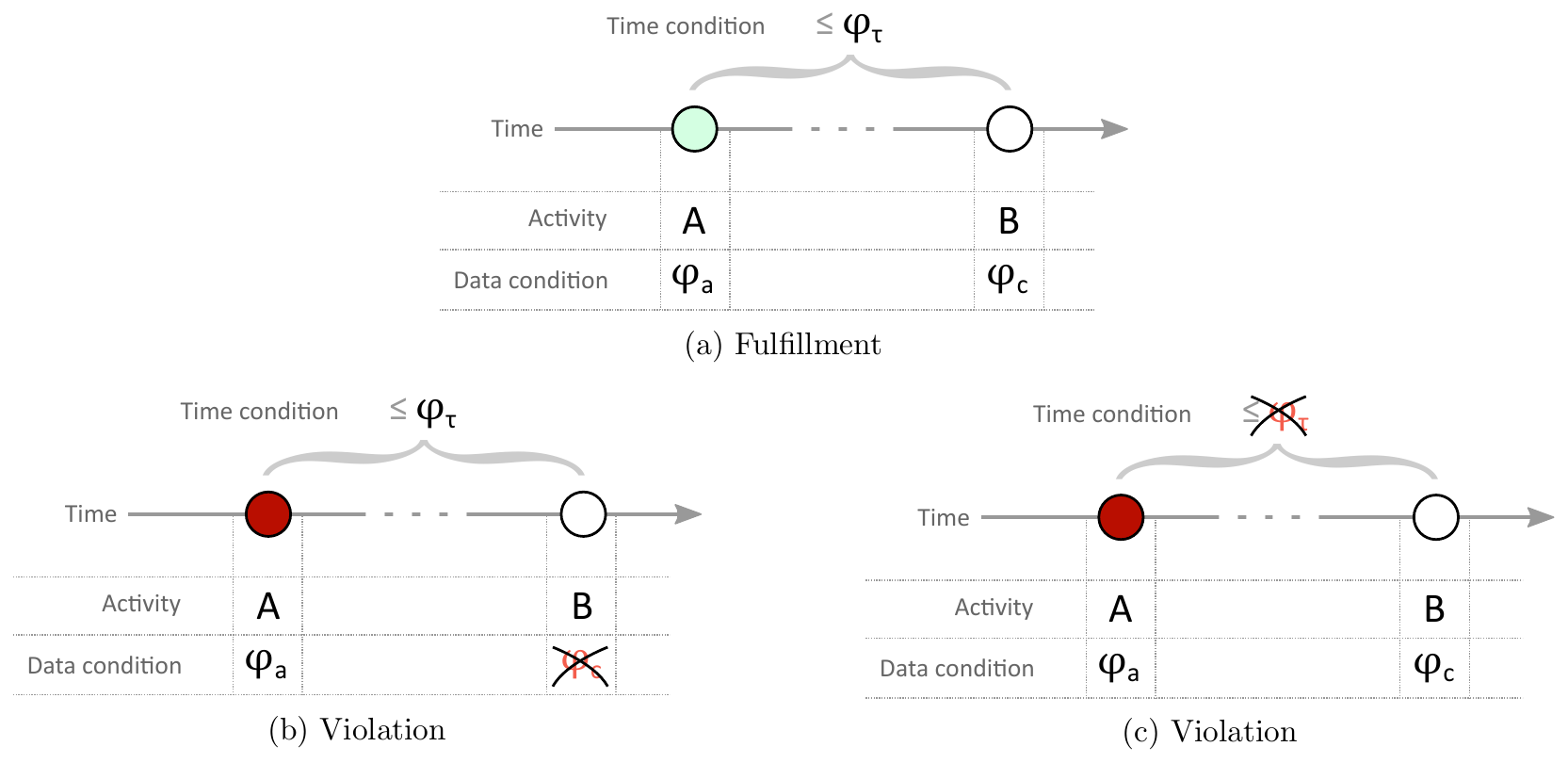}
	\caption{Fulfillment and violation scenarios for the \emph{response} constraint between activities $A$ and $B$. (a) reports a typical fulfillment scenario. In (b), the violation is due to the violation of the correlation condition $\varphi_c$. In (c), the violation is due to the violation of the time condition $\varphi_{\tau}$.}
	\label{fig:template:response}
\end{figure}

\begin{figure}[t!]
	\includegraphics[width=\textwidth]{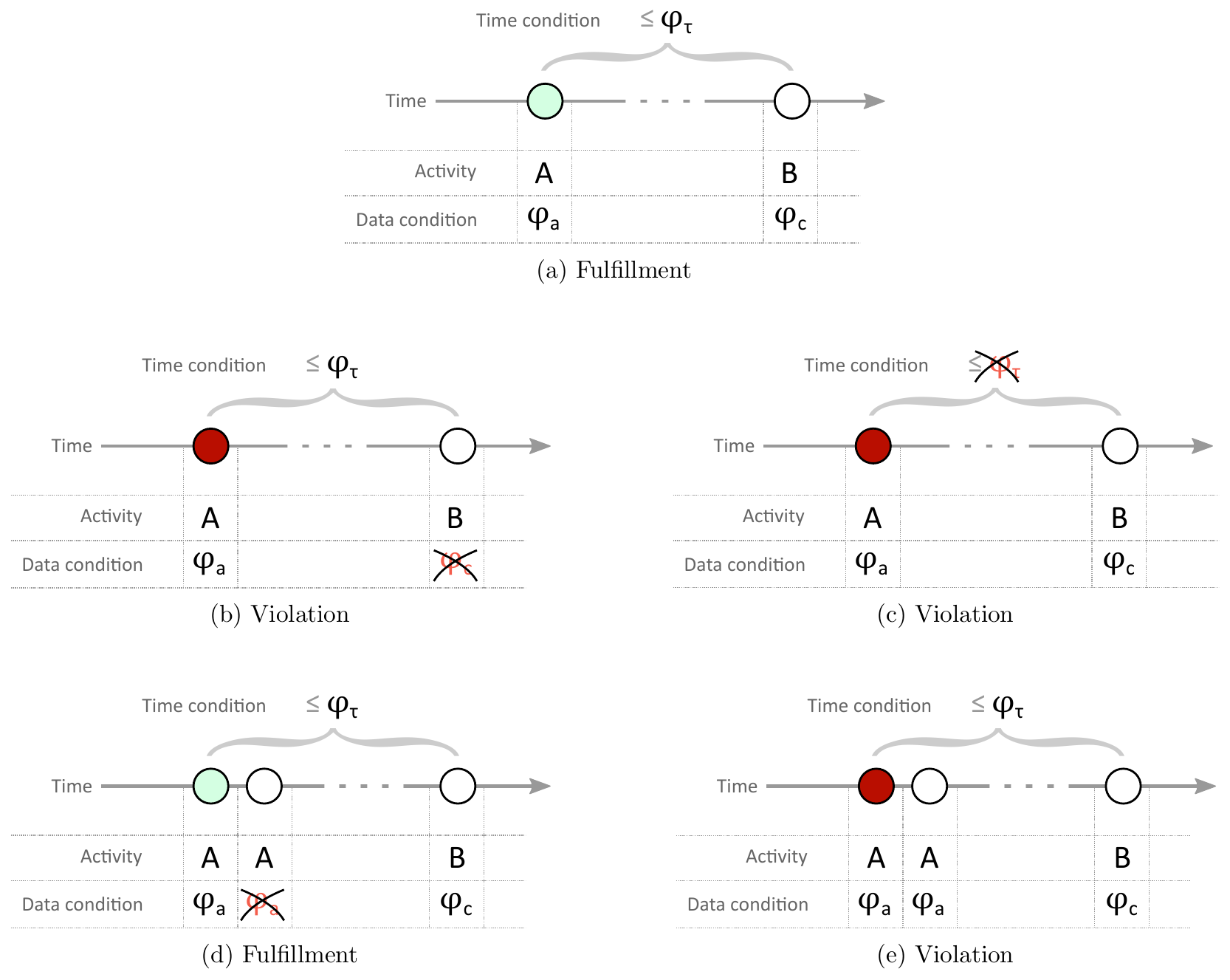}
	\caption{Fulfillment and violation scenarios for the \emph{alternate response} constraint between activities $A$ and $B$. (a) reports a typical fulfillment scenario. In (b), the violation is due to the violation of the correlation condition $\varphi_c$. In (c), the violation is due to the violation of the time condition $\varphi_{\tau}$. The activation in (d) is a fulfillment because the second occurrence of $A$ does not satisfy the activation condition. In contrast, (e) reports a violation since, in this case, the second occurrence of $A$ satisfies the activation condition.}
	\label{fig:template:alternateresponse}
\end{figure}

\begin{figure}[t!]
	\includegraphics[width=\textwidth]{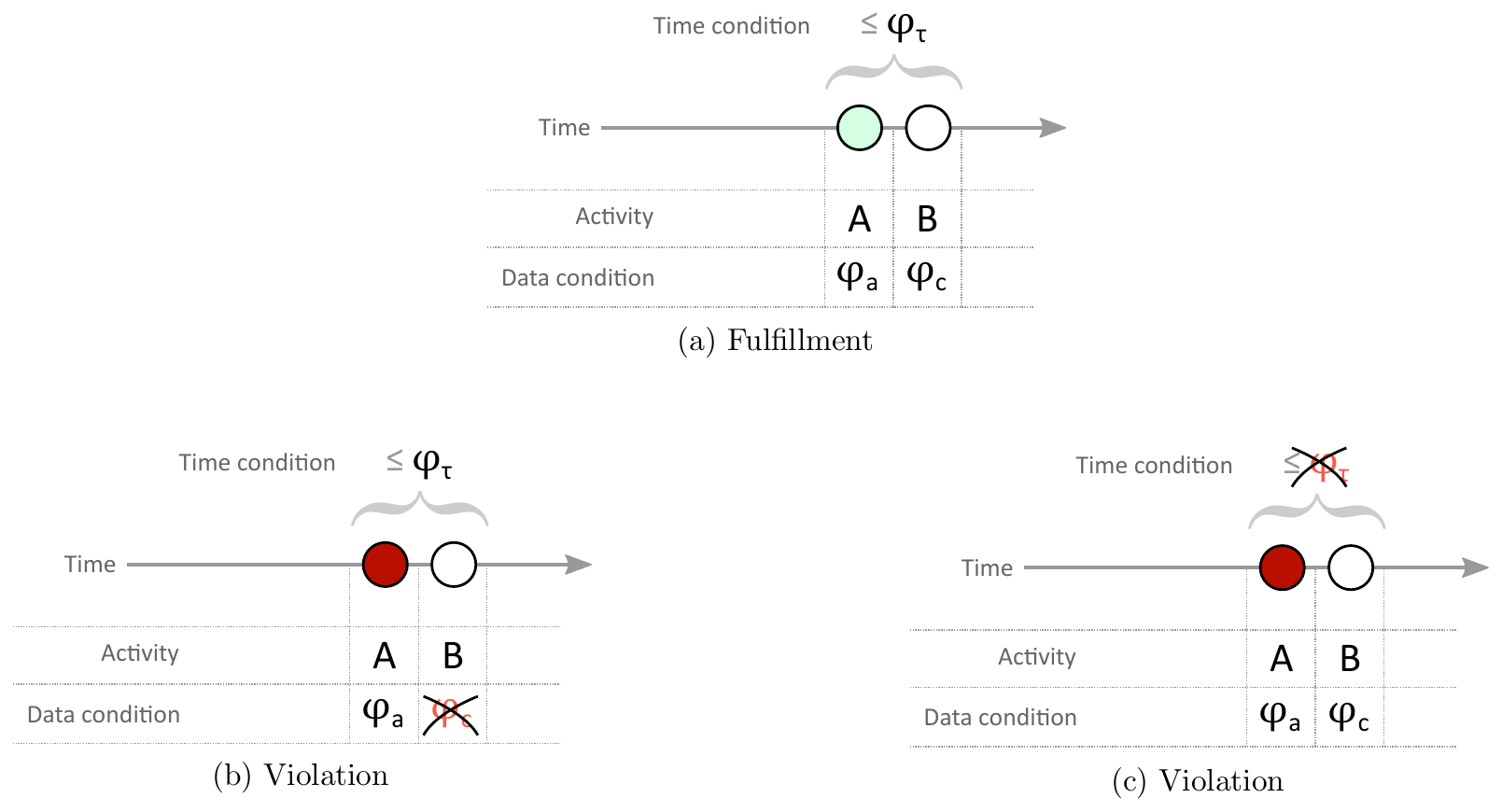}
	\caption{Fulfillment and violation scenarios for the \emph{chain response} template between activities $A$ and $B$. (a) reports a typical fulfillment scenario. Note that, in this case, the two events are contiguous. In (b), the violation is due to the violation of the correlation condition $\varphi_C$. In (c), the violation is due to the violation of the time condition $\varphi_{\tau}$.}
	\label{fig:template:chainresponse}
\end{figure}
\section{MFOTL Semantics for Multi-Perspective Business Constraints}
\label{sec:semantics}
In this section, we introduce a multi-perspective version of Declare (\MPDeclare).  The version is
similar to the ones in \cite{timeddeclare,DBLP:conf/ispw/MasellisMM14}, but we enrich it by allowing both time and data perspective. To do this, we use Metric First-Order Linear Temporal Logic (MFOTL). While many reasoning
tasks are clearly undecidable for MFOTL, this logic is appropriate to unambiguously
describe the semantics of the \MPDeclare~constraints we can use for conformance checking in our proposed algorithms.

To define the new semantics for Declare, we have to contextualize the definitions given in Section \ref{sec:mfotl} in XES. Consider, for example, that the execution of an activity $pay$ is recorded in an event log and, after the execution of $pay$ at timestamp $\tau_i$, the attributes $originator$, $amount$, and $z$ have values $John$, $100$, and $July$. In this case, the valuation of $(activityName,originator,amount,z)$ is $(pay,John,100,July)$ in $\tau_i$. Considering that in XES, by definition, the activity name is a special attribute always available, if $(pay,John,100,July)$ is the valuation of $(activityName,originator,amount,z)$, we say that, when $pay$ occurs, two special relations are valid $event(pay)$ and $p_{pay}(John,100,July)$. In the following, we identify $event(pay)$ with the event itself $pay$ and we call $(John,100,July)$, the \emph{payload} of $pay$.

The semantics for \MPDeclare ~is shown in Table~\ref{tbl:timed-mfotl}. Note that all the templates here considered have two parameters, an activation and a target (see also \tablename~\ref{tbl:timed-ltl}). As an example, we consider the response constraint ``activity \emph{pay} is always eventually followed by activity \emph{get discount}'' having \emph{pay} as activation and \emph{get discount} as target. The timed semantics of Declare, introduced in \cite{timeddeclare}, is extended by requiring two additional conditions on data, i.e., the \emph{activation condition} $\varphi_a$ and the \emph{correlation condition} $\varphi_c$.
The activation condition is a relation (over the variables corresponding to the global attributes in the event log) that must be valid when the activation occurs. If the activation condition does not hold the constraint is not activated. In the case of the response template the activation condition has the form $p_A(x) \wedge r_{a}(x)$, meaning that when $A$ occurs with payload $x$, the relation $r_a$ over $x$ must hold. For example, we can say that whenever \emph{pay} occurs and \emph{client type} is \emph{gold} then eventually \emph{get discount} must follow. In case \emph{pay} occurs but \emph{client type} is not \emph{gold} the constraint is not activated.
The correlation condition is a relation that must be valid when the target occurs. It has the form $p_B(y) \wedge r_{c}(x,y)$, where $r_c$ is a relation involving, again, variables corresponding to the (global) attributes in the event log but, in this case, relating the valuation of the attributes corresponding to the payload of $A$ and the valuation of the attributes corresponding to the payload of $B$. In our example, we can say that whenever \emph{pay} occurs and \emph{client type} is \emph{gold} then eventually \emph{get discount} must follow and the due amount corresponding to activity \emph{get discount} must be lower than the one corresponding to activity \emph{pay}. In the following, with abuse of notation we specify the interval characterizing the time perspective of a \MPDeclare~constraint ($I=[a,b)$) with $\varphi_{\tau}$.

\begin{table}
\caption{Semantics for \MPDeclare\ constraints. \label{tbl:timed-mfotl}}
\centering
\scriptsize{
\begin{tabular}{ll}
\toprule
\textbf{Template} & \textbf{MFOTL Semantics} \\
\midrule
responded existence  & $\lglobally( \forall x.((A \wedge \varphi_a(x)) \rightarrow (\tonce (B   \wedge \exists y.\varphi_c(x,y)) \vee \tfuture (B \wedge \exists y.\varphi_c(x,y)))))$ \\
\midrule
response &  $\lglobally( \forall x. ((A \wedge \varphi_a(x)) \rightarrow \tfuture (B \wedge \exists y.\varphi_c(x,y))))$ \\
alternate response  & $ \lglobally(\forall x. ((A \wedge \varphi_a(x)) \rightarrow \lnext(\neg (A \wedge \varphi_a(x)) \tuntil (B \wedge \exists y.\varphi_c(x,y)))))$ \\
chain response &  $\lglobally(\forall x. ((A \wedge \varphi_a(x)) \rightarrow \tnext (B \wedge \exists y.\varphi_c(x,y)))$ \\
\midrule
precedence &  $\lglobally(\forall x. ((B \wedge \varphi_a(x)) \rightarrow \tonce (A \wedge \exists y.\varphi_c(x,y)))$ \\
alternate precedence & $ \lglobally(\forall x. ((B \wedge \varphi_a(x)) \rightarrow \lyesterday(\neg (B \wedge \varphi_a(x)) \tsince (A \wedge \exists y.\varphi_c(x,y))))$ \\
chain precedence & $\lglobally(\forall x. ((B \wedge \varphi_a(x)) \rightarrow \tyesterday (A \wedge \exists y.\varphi_c(x,y)))$ \\
\midrule
not responded existence  &
$\lglobally( \forall x.((A \wedge \varphi_a(x)) \rightarrow \neg (\tonce (B   \wedge \exists y.\varphi_c(x,y)) \vee \tfuture (B \wedge \exists y.\varphi_c(x,y)))))$ \\
not response  & $\lglobally( \forall x. ((A \wedge \varphi_a(x)) \rightarrow \neg \tfuture (B \wedge \exists y.\varphi_c(x,y))))$ \\
not precedence & $\lglobally(\forall x. ((B \wedge \varphi_a(x)) \rightarrow \neg \tonce (A \wedge \exists y.\varphi_c(x,y)))$ \\
not chain response  & $\lglobally(\forall x. ((A \wedge \varphi_a(x)) \rightarrow \neg \tnext (B \wedge \exists y.\varphi_c(x,y)))$ \\
not chain precedence  & $\lglobally(\forall x. ((B \wedge \varphi_a(x)) \rightarrow \neg \tyesterday (A \wedge \exists y.\varphi_c(x,y)))$ \\
\bottomrule
\end{tabular}}
\end{table}

Graphical representations of three \MPDeclare~templates are reported in Figures~\ref{fig:template:response}, \ref{fig:template:alternateresponse} and \ref{fig:template:chainresponse}. In particular, these figures report the semantics for response, alternate response and chain response constraints. Each figure shows possible scenarios of violations and fulfillments for the corresponding constraint. A scenario is described reporting events as rounded circles. Each circle is associated to an activity ($A$, $B$, or $C$) and a data condition (either an activation condition $\varphi_a$ or a correlation condition $\varphi_c$). The time condition $\varphi_{\tau}$ is reported above the horizontal curly bracket. Crossed data or time conditions indicate violated conditions. Red circles indicate events that are violations, green circles indicate fulfillments.

The \emph{response} constraint in Figure~\ref{fig:template:response} indicates that, if $A$ occurs at time $\tau_A$ with $\varphi_a$ holding true, $B$ must occur at some point $\tau_B\in[\tau_A + a, \tau_A + b)$ with $\varphi_c$ holding true.
The \emph{alternate response} constraint in Figure~\ref{fig:template:alternateresponse} specifies that,
if $A$ occurs at time $\tau_A$ with $\varphi_a$ holding true, $B$ must occur at some point $\tau_B\in[\tau_A + a, \tau_A + b)$ with $\varphi_c$ holding true. $A$ is not allowed in the interval $[\tau_A,\tau_B]$ if $\varphi_a$ is true. Any event different from $A$ is allowed and, also, $A$ is allowed if $\varphi_a$ is false.
The \emph{chain response} constraint in Figure~\ref{fig:template:chainresponse} indicates that, if $A$ occurs at time $\tau_A$ with $\varphi_a$ holding true, $B$ must occur next at some point $\tau_B\in[\tau_A + a, \tau_A + b)$ with $\varphi_c$ holding true.

%% file: sections/algorithms.tex
\section{Conformance Checking Algorithms}
\label{sec:algorithms}


As stated in the previous section, with \MPDeclare{}, it is possible to express Declare constraints taking into account also the temporal and the data perspectives. As an example, it is possible to express constraints like:
\begin{itemize}
	\item \textit{activity $A$ must occur between 10 and 11 hours before activity $B$};
	\item \textit{if activity $A$ writes a variable $x$ with value $<$1000, then $B$ must occur after two days}.
\end{itemize}
Therefore, using this language, it is possible to define multi-perspective compliance models that can be used for several purposes like, for example, for representing Service Level Agreements (SLAs). In this context, it would be useful to provide the user with techniques to detect whether cases are actually fulfilling the required set of constraints or not. In this section, we present algorithms to check the conformance of an event log with respect to a \MPDeclare\ model.

The proposed approach for the conformance checking of \MPDeclare\ constraints is based on several procedures. The main component is described in the \texttt{CheckLogConformance} procedure, reported in Algorithm~\ref{alg:conformance-log}. This algorithm requires as input a log and a \MPDeclare\ model (i.e., a set of \MPDeclare\ constraints). Then, it iterates through all traces and, for each constraint, it computes the violations and the fulfillments by calling the \texttt{CheckTraceConformance} procedure.
\texttt{CheckTraceConformance}, described in Algorithm~\ref{alg:conformance-trace}, takes as input a trace and a constraint and generates the set of violating and fulfilling events for that specific constraint in that specific trace. The basic idea of this procedure is to iterate through all the events of the trace and, for each of them, call specific template-dependent operations (lines~5-11).
%
%
\begin{algorithm2e}[h]
	\DontPrintSemicolon
	\KwIn{$\textit{Log}$: an event log}
	\KwInXX{$\textit{Model}$: a model}
	\KwOut{A set of violating and fulfilling traces/constraints}
	\BlankLine
	Let $\textit{fulfill}$ and $\textit{viol}$ be maps that, given a trace and a constraint, return the set of fulfilling and violating events \;
	\BlankLine
	\ForEach{$\textit{trace} \in \textit{Log}$} {
		\ForEach{$\textit{constr} \in \textit{Model}$} {
			$\textit{viol}, \textit{fulfill} \gets \texttt{CheckTraceConformance}(\textit{trace},\textit{constr})$ \tcp*[r]{Algorithm~\ref{alg:conformance-trace}}
			\BlankLine
			$\textit{viol}\ [\textit{trace}][\textit{constr}] \gets \textit{viol}$ \;
			$\textit{fulfill}\ [\textit{trace}][\textit{constr}] \gets \textit{fulfill}$ \;
		}
	}
	\Return{$\textit{viol}, \textit{fulfill}$}
	\caption{\texttt{CheckLogConformance}}
	\label{alg:conformance-log}
\end{algorithm2e}

\begin{algorithm2e}[h]
	\DontPrintSemicolon
	\KwIn{$\textit{trace}$: a trace}
	\KwInXX{$c = \langle \textit{templ}, A, T, \varphi_a, \varphi_c, \varphi_{\tau} \rangle$: a  constraint}
	\KwOut{Set of violating and fulfilling events}
	\BlankLine
	$\textit{pending} \gets \emptyset$ \;
	$\textit{fulfillments} \gets \emptyset$ \;
	$\textit{violations} \gets \emptyset$ \;
	\BlankLine
	\tcc{All the following calls are allowed to make side effects on the provided parameters}
	$\textit{templ}.\textit{opening}()$ \tcc*{Opening template operations}
	\ForEach{$e \in \textit{trace}$} {
		$\textit{templ}.\textit{fulfillment}(e, \textit{trace}, \textit{pending}, \textit{fulfillments}, T, \varphi_a, \varphi_c, \varphi_{\tau})$ \; \label{alg:conformance-trace:start-for}
		$\textit{templ}.\textit{violation}(e, \textit{trace}, \textit{pending}, \textit{violations}, T, \varphi_c, \varphi_{\tau})$ \; \label{alg:conformance-trace:end-for}
		$\textit{templ}.\textit{activation}(e, A, \textit{pending}, \varphi_a)$ \;
	}
	$\textit{templ}.\textit{closing}(\textit{pending}, \textit{fulfillments}, \textit{violations})$ \tcc*{Closing template operations}
	\BlankLine
	\Return{$\textit{violation}, \textit{fulfillments}$}
	\caption{\texttt{CheckTraceConformance}}
	\label{alg:conformance-trace}
\end{algorithm2e}

The described algorithms might be seen as a general ``framework'' that can be used for conformance checking with respect to different templates. Each template that needs to be verified must properly define the following required operations:
\begin{itemize}
	\item \emph{opening}: this procedure is called once per trace, before starting the analysis of the first event of the trace;
	\item \emph{fulfillments}: this procedure is called for each event of the trace and is supposed to return the set of fulfillments that have been observed so far; modifications to the set of activations are allowed as well;
	\item \emph{violations}: this procedure is called for each event of the trace and is supposed to return the set of violations that have been observed so far; modifications to the set of activations are allowed as well;
    \item \emph{activation}: this procedure is called for each event of the trace and is supposed to update the set of activations that have been observed so far (i.e., whether the current event is a new activation or not);
	\item \emph{closing}: this procedure is called once per trace, after all the events have been analyzed.
\end{itemize}

In this paper, we illustrate the procedures for three templates, i.e., \emph{response}, \emph{alternate response}, and \emph{chain response}. We consider these three specifications sufficiently representative in order to provide a clear idea of the capabilities of our framework.\footnote{All the procedures for conformance checking based on \MPDeclare{} have been implemented and are publicly available (see Section \ref{sec:benchmarks}).}
In each procedure, given the set of all possible activities $\mathcal{A}$, we define a constraint as a tuple:
$c = \langle \textit{template}, A, T, \varphi_a, \varphi_c, \varphi_{\tau} \rangle$, where $\textit{template}$ indicates which template the constraint is referring to, $\textit{template} \in \{\textit{existence}, \textit{absence}, \textit{choice}, \textit{responded existence},\dots \}$;
$A \subseteq \mathcal{A}$ is the nonempty set of activations; $T \subseteq \mathcal{A}$ is the nonempty set of targets; $\varphi_a$ and $\varphi_c$ indicate, respectively, the activation and the correlation condition; and $\varphi_{\tau}$ represents the time condition. We also use the functions $\verify(\varphi_a, A)$, $\verify(\varphi_c, A, B)$, and $\verify(\varphi_{\tau}, A, B)$.
The first function evaluates $\varphi_a$ with respect to the attributes reported in $A$. The second function evaluates $\varphi_c$ with respect to the attributes defined in $A$ and $B$. The third function compares the timestamps attached to $A$ and $B$ in order to see whether $\varphi_{\tau}$ is satisfied or not. As already mentioned, each event recorded in an event log brings a payload of attributes. In the description of the algorithms, we use the $\pi_a(e)$ operator to get the value of an attribute $a$ of an event $e$. 
For example, we use $\pi_\textit{activity}(e)$ to select the activity name associated to $e$. 


The first template we consider is \emph{response} and the corresponding procedures are reported in Table~\ref{tbl:constraint:response}.
The \emph{opening} procedure does nothing. The \emph{fulfillment} procedure checks whether the input event refers to a target. If this is the case, then all pending activations that can be correlated to this target (in case the time and the correlation conditions are satisfied) become fulfillments.
The \emph{activation} procedure checks whether the input event refers to an activation of the constraint and the activation condition $\varphi_a$ is satisfied (in this case the event has to be added to the set of pending activations).
Violations are identified in the \emph{closing} procedure (the \emph{violation} procedure is not used in this case). Here, all pending activations that do not have a corresponding target when the entire trace has been processed become violations.

\templateDefinition
{Response} 
{Procedure specifications for the \textit{response} constraint.} 
{tbl:constraint:response} 
{ 
	\textbf{do nothing} \;
} 
{ 
	\If{$\pi_\textit{activity}(e) \in A$ \textbf{\emph{and}} $\verify(\varphi_a, e)$} {
		$\textit{pending} \gets \textit{pending} \cup \{ e \}$ \;
	}
} 
{ 
	\If{$\pi_\textit{activity}(e) \in T$} {
		\ForEach{$\textit{act} \in \textit{pending}$} {
			\If{$\verify(\varphi_c, \textit{act}, e)$ \textbf{\emph{and}} $\verify(\varphi_{\tau}, \textit{act}, e)$} {
				$\textit{pending} \gets \textit{pending} \setminus \{ \textit{act} \}$ \;
				$\textit{fulfillments} \gets \textit{fulfillments} \cup \{ \textit{act} \}$ \;
			}
		}
	}
} 
{ 
	\textbf{do nothing} \tcc*{Actual violations are not identified here}
} 
{ 
	\ForEach{$\textit{act} \in \textit{pending}$}{
		$\textit{pending} \gets \textit{pending} \setminus \{ \textit{act} \}$ \;
		$\textit{violations} \gets \textit{violations} \cup \{ \textit{act} \}$ \;
	}
} 


The procedures for the \emph{alternate response} template are reported in Table~\ref{tbl:constraint:alternateresponse}. In particular, \emph{opening} defines a new data structure ($\textit{possibleTargets}$) that will be used by the other procedures.
The \emph{fulfillment} procedure starts by checking whether the input event refers to an activation and the activation condition is satisfied. If this is the case, the procedure checks whether there is exactly one pending activation and at least one possible target. If this is the case, if for at least one possible target the time and the correlation conditions are satisfied, the pending activation becomes a fulfillment (\emph{fulfillment}, lines~\ref{tbl:constraint:alternateresponse:fulfill}-8). If the activity referring to the input event is a target, the event is added to the set of possible targets (\emph{fulfillment}, line~\ref{tbl:constraint:alternateresponse:targets}).
The \emph{violation} procedure also starts by checking whether the input event refers to an activation and the activation condition is satisfied. If this is the case, the procedure checks whether there is exactly one pending activation.
If this is the case,
the pending activation becomes a violation (the pending activation cannot be a fulfillment because, in this case, the invocation of the \emph{fulfillment} procedure moves it from the pending set to the fulfillment set).
%
The \emph{activation} procedure checks whether the input event refers to an activation and the activation condition is satisfied. In this case, the set of possible targets is reset to the empty value and the event is returned to be added to the set of pending activations.
The \emph{closing} procedure verifies that if there is a pending activation, this activation can be correlated at least to one possible target. If this is the case (if the time and the correlation conditions are satisfied), then the activation becomes a fulfillment (\emph{closing}, line~\ref{tbl:constraint:alternateresponse:fulfill2}), otherwise it is marked as a violation (\emph{closing}, line~\ref{tbl:constraint:alternateresponse:newviolation3}).

\templateDefinition
{Alternate Response} 
{Procedure specifications for the \textit{alternate response} constraint.} 
{tbl:constraint:alternateresponse} 
{ 
	\textbf{define} $\textit{possibleTargets} \gets \emptyset$ as a data structure available throughout the entire \texttt{CheckTraceConformance} algorithm \;
} 
{ 
	\If{$\pi_\textit{activity}(e) \in A$ \textbf{\emph{and}} $\verify(\varphi_a, e)$} {
		$\textit{possibleTargets} \gets \emptyset$ \;
		$\textit{pending} \gets \textit{pending} \cup \{ e \}$ \;
	}
} 
{ 
	\If{$\pi_\textit{activity}(e) \in A$ \textbf{\emph{and}} $\verify(\varphi_a, e)$} {
		\If{$|\textit{possibleTargets}| \geq 1$ \textbf{\emph{and}} $|\textit{pending}| = 1$} {
			$\textit{act} \gets \textit{element} \in \textit{pending}$ \tcp{There is only one element}
			\ForEach{$p \in \textit{possibleTargets}$} {
				\If{$\verify(\varphi_c, \textit{act}, p)$ \textbf{\emph{and}} $\verify(\varphi_{\tau}, \textit{act}, p)$} {
					$\textit{fulfillments} \gets \textit{fulfillments} \cup \{ \textit{act} \}$ \; \label{tbl:constraint:alternateresponse:fulfill}
					$\textit{pending} \gets \textit{pending} \setminus \{ \textit{act} \}$ \;
					\textbf{break} \tcp{It is possible to exit the loop}
				}
			}
		}
	}
	\If{$e \in T$} {
		$\textit{possibleTargets} \gets \textit{possibleTargets} \cup \{ e \}$ \; \label{tbl:constraint:alternateresponse:targets}
	}
} 
{ 
	\If{$\pi_\textit{activity}(e) \in A$ \textbf{\emph{and}} $\verify(\varphi_a, e)$} {
		\If{$|\textit{pending}| = 1$} {
			$\textit{act} \gets \textit{element} \in \textit{pending}$ \tcp{There is only one element}
			$\textit{pending} \gets \textit{pending} \setminus \{ \textit{act} \}$ \;
			$\textit{violations} \gets \textit{violations} \cup \{ \textit{act} \}$ \; \label{tbl:constraint:alternateresponse:newviolation2}
		}
	}
} 
{ 
	\If{$|\textit{pending}| = 1$} {
		$\textit{targetFound} \gets$ false \;
		$\textit{act} \gets \textit{element} \in \textit{pending}$ \tcp{There is only one element}
		\ForEach{$p \in \textit{possibleTargets}$} {
			\If{	$\verify(\varphi_c, \textit{act}, p)$ \textbf{\emph{and}} $\verify(\varphi_{\tau}, \textit{act}, p)$} {
				$\textit{targetFound} \gets$ true \;
				$\textit{fulfillments} \gets \textit{fulfillments} \cup \{ \textit{act} \}$ \; \label{tbl:constraint:alternateresponse:fulfill2}
			}
		}
		\If{\emph{\textbf{not}} $\textit{targetFound}$} {
			$\textit{violations} \gets \textit{violations} \cup \{ \textit{act} \}$ \; \label{tbl:constraint:alternateresponse:newviolation3}
		}
	}
} 


The procedures for the \emph{chain response} template are reported in Table~\ref{tbl:constraint:chainresponse}. As for the response template, \emph{opening} does nothing.
The \emph{fulfillment} and the \emph{violation} procedures verify whether there is exactly one element in the set of pending activations. In this case, they check whether the input event refers to a target and the time and correlation conditions are fulfilled. If this is the case, the pending activation becomes a fulfillment, otherwise it is marked as a violation.
The \emph{activation} procedure checks whether the input event refers to an activation and the activation condition is satisfied (in this case the event has to be added to the set of pending activations).
The \emph{closing} procedure checks whether there is still a pending activation when the entire trace has been processed. In this case, the pending activation becomes a violation.

\templateDefinition
{Chain Response} 
{Procedure specifications for the \textit{chain response} constraint.} 
{tbl:constraint:chainresponse} 
{ 
	\textbf{do nothing} \;
} 
{ 
	\If{$\pi_\textit{activity}(e) \in A$ \textbf{\emph{and}} $\verify(\varphi_a, e)$} {
		$\textit{pending} \gets \textit{pending} \cup \{ e \}$ \;
	}
} 
{ 
	\If{$|\textit{pending}| = 1$} {
		$\textit{act} \gets \textit{element} \in \textit{pending}$ \tcp{There is only one element}
		\If{$\pi_\textit{activity}(e) \in T$ \textbf{\emph{and}}
			$\verify(\varphi_c, \textit{act}, e)$ \textbf{\emph{and}}
			$\verify(\varphi_{\tau}, \textit{act}, e)$} {
			$\textit{pending} \gets \textit{pending} \setminus \{ \textit{act} \}$ \;
			$\textit{fulfillments} \gets \textit{fulfillments} \cup \{ \textit{act} \}$ \;
		}
	}
} 
{ 
	\If{$|\textit{pending}| = 1$} {
		$\textit{act} \gets \textit{element} \in \textit{pending}$ \tcp{There is only one element}
		\If{$\pi_\textit{activity}(e) \notin T$ \textbf{\emph{or not}}
			$\verify(\varphi_c, \textit{act}, e)$ \textbf{\emph{or not}}
			$\verify(\varphi_{\tau}, \textit{act}, e)$} {
			$\textit{pending} \gets \textit{pending} \setminus \{ \textit{act} \}$ \;
			$\textit{violations} \gets \textit{violations} \cup \{ \textit{act} \}$ \;
		}
	}
} 
{ 
	\ForEach{$\textit{act} \in \textit{pending}$}{
		$\textit{pending} \gets \textit{pending} \setminus \{ \textit{act} \}$ \;
		$\textit{violations} \gets \textit{violations} \cup \{ \textit{act} \}$ \;
	}
} 

The algorithms for the other templates specified in Table~\ref{tbl:timed-mfotl} can be very easily derived from the ones described in this section. In particular, the algorithms for the \emph{precedence}, the \emph{alternate precedence} and the \emph{chain precedence} are the same as the ones described for \emph{response}, \emph{alternate response} and \emph{chain response} respectively. The only difference is that, for the precedence templates, the traces in the input log have to be parsed from the end to the beginning. Similarly, the algorithms for checking the negative templates are the same as the ones described for the corresponding negative templates. In this case, every fulfillment for a positive template becomes a violation for the corresponding negative template and vice versa.

From the computational complexity point of view, it is worthwhile noting that the complexity of Algorithm~\ref{alg:conformance-log} and Algorithm~\ref{alg:conformance-trace} is linear in the number of traces, constraints, and in the number of events of each trace. The complexity of the template-dependent procedures, instead, depends on the actual template. Specifically, with respect to the procedures of each constraint reported in this paper, we have the following complexities:
\begin{itemize}
	\item Response: \textit{opening}, \textit{violation}, and \textit{activation} are constant; \textit{fulfillment} and \textit{closing} have linear complexity on the number of pending activations for the current trace (which is at most the number of events on the trace);
	\item Alternate Response: \textit{opening}, \textit{violation}, and \textit{activation} are constant; \textit{fulfillment} and \textit{closing} are linear on the number of possible targets (which is at most the number of events on the trace);
	\item Chain Response: \textit{opening}, \textit{fulfillment}, \textit{violation} , \textit{activation} are constant; \textit{closing} is linear on the number of pending activations for the current trace (which is at most the number of events on the trace).
\end{itemize}

%% file: sections/benchmarks.tex
\section{Implementation and Benchmarks}
\label{sec:benchmarks}

This section provides some details on the implementation of the approach and a benchmark analysis on different scenarios.

\subsection{Implementation Details}

The entire approach has been implemented as a plug-in of the process mining toolkit ProM.\footnote{The software can be downloaded from \url{http://www.promtools.org/prom6}.} In particular, the plug-in receives as input an event log and a model and evaluates the conformance of the log with respect to the model.
It is interesting to note that, in the current implementation, the processing of each trace is independent from all the others. Also, the analysis of a constraint in the reference model is independent from all the others. For this reason, it is possible to parallelize and distribute the analysis over different computational nodes and drastically improve the performances. The results of the tests reported in this paper, however, do not benefit from such a possibility and our tests sequentially evaluate each constraint on each trace.

\begin{figure}
	\centering
	 \includegraphics[width=.8\textwidth]{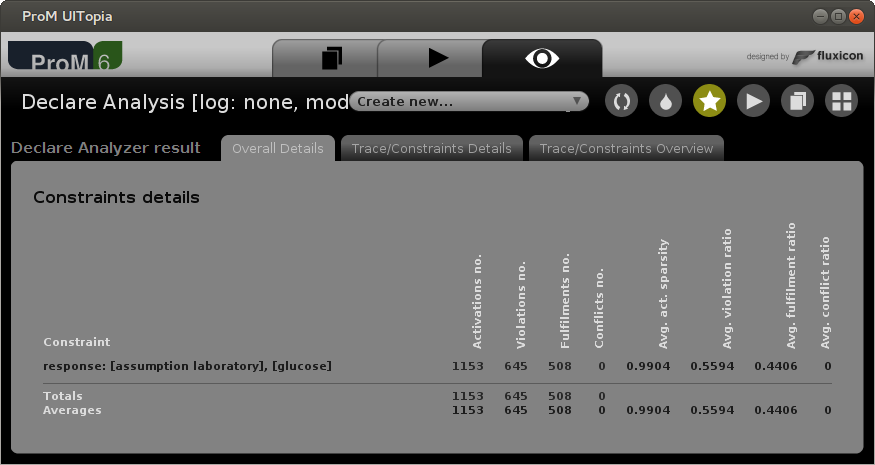}
	\caption{Overall details window with the result summary.}
	\label{fig:screenshot:overall}
\end{figure}

The conformance checking results are presented using a ProM plug-in called ``Analysis Result Visualizer''. This visualizer is composed of three main windows. The first window consists of a summary of the statistics computed for each constraints (e.g., number of activations, number of violations and number of fulfillments) on the entire log. This window is shown in \figurename~\ref{fig:screenshot:overall}.

\begin{figure}
	\centering
	\begin{subfigure}{\textwidth}
		\includegraphics[width=\textwidth]{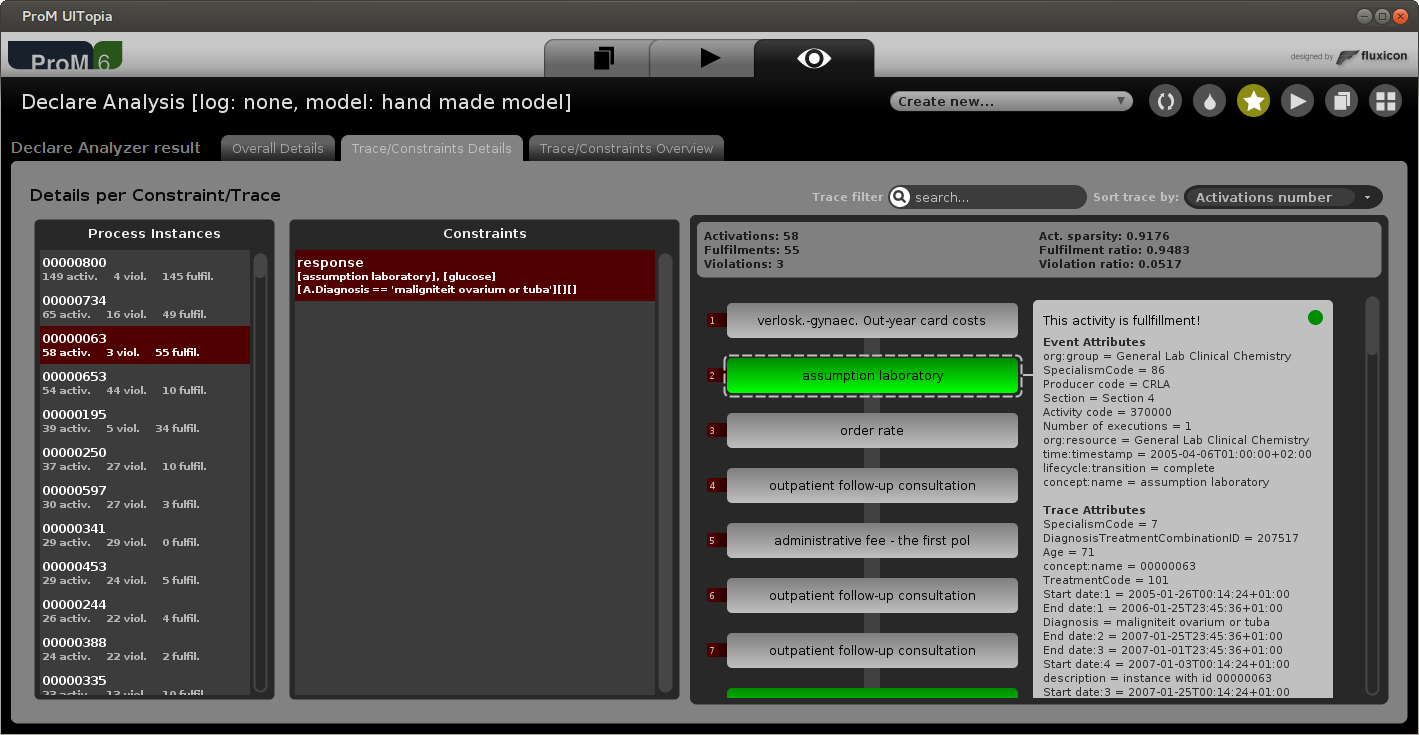}
		\caption{Window with the conformance checking details for a single case and constraint.}
		\label{fig:screenshot:newdetails}
	\end{subfigure}
	\begin{subfigure}{\textwidth}
		\includegraphics[width=\textwidth]{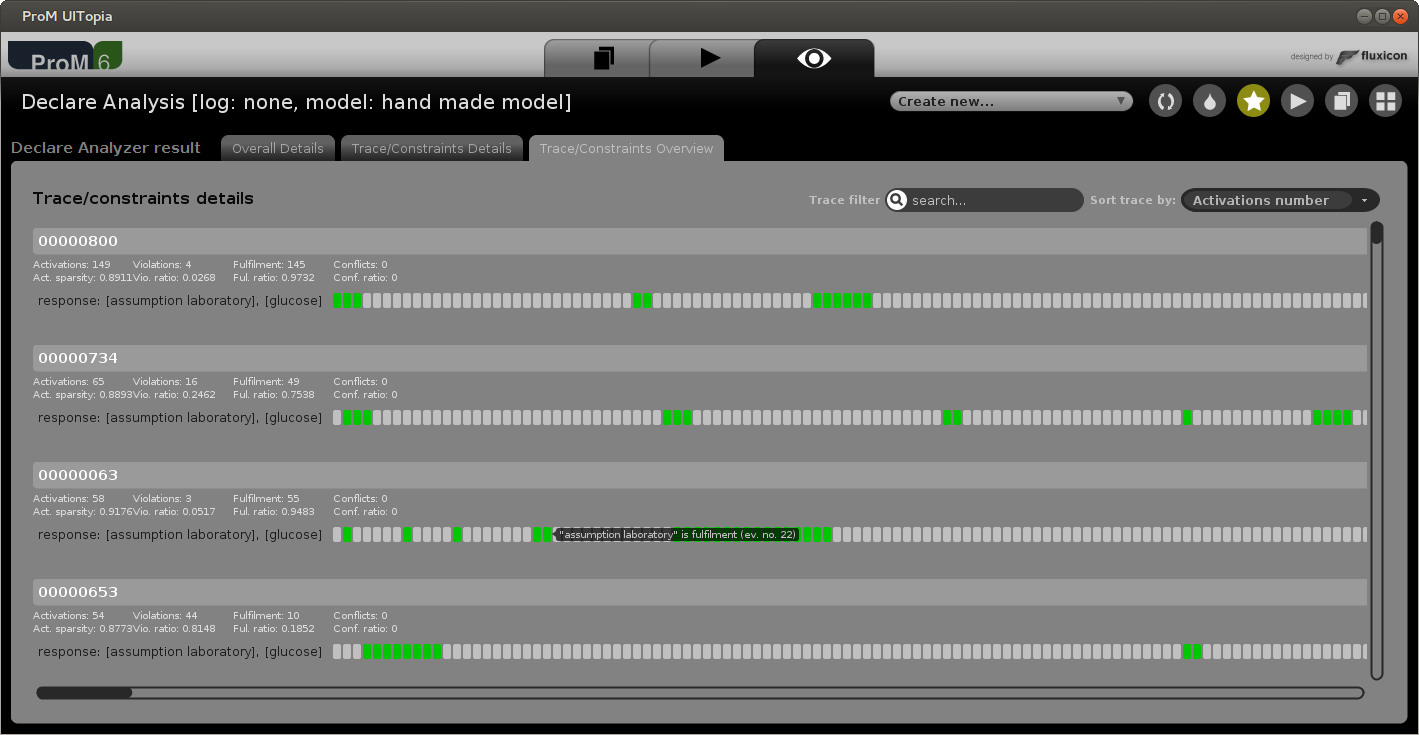}
		\caption{Birdview-like window showing an overview of fulfillments and violations for some cases.}
		\label{fig:screenshot:details}
	\end{subfigure}
	\caption{Windows used to inspect the conformance checking results by focusing on single cases.}
	\label{fig:screenshot:details:overall}
\end{figure}
The second window (shown in \figurename~\ref{fig:screenshot:newdetails}) provides a more detailed view. This window is divided into three columns. The leftmost column contains a list of all the cases with information on case id, number of activations in the case, and number of fulfillments and violations. The central column contains the list of constraints in the reference model. When a case and a constraint are selected, in the list in the rightmost column of the window, a representation of the case appears. In this representation, each event is drawn as a rectangle. Green-painted rectangles represent fulfillments, red-painted boxes represent violations. It is possible to move the mouse cursor over each rectangle to see the complete set of attributes belonging to the event.

The third window (shown in \figurename~\ref{fig:screenshot:details}) also lists all cases. Here, each event of a case is represented as a small box that can be gray, green (in case the event is a fulfillment), or red (in case the event is a violation). This visualization is also called ``birdview'' since it provides a high-level overview of the constraints and allows the user to quickly identify possible issues. When the mouse is moved over an event, a pop-up showing the corresponding activity name appears.
In both the second and the third window, it is possible to sort cases based on different parameters (name of the case, number of activations, number of violations, and number of fulfillments), or interactively search for cases with a specific case id.

\subsection{Benchmarks}

In order to gain some insights on the computational feasibility of our implementation, we run several tests in different possible scenarios. In particular, we tested our implementation against logs with different sizes and different trace lengths. We generated traces with 10, 20, 30, 40, and 50 events and, for each of these lengths, we generated logs with 25\,000, 50\,000, 75\,000, and 100\,000 traces. Therefore, in total, we used 20 logs. The number of events contained in each log is reported in \tablename~\ref{tbl:events-per-log}. In addition, we designed 10 Declare models. In particular, we prepared two models with 10 constraints, one only containing constraints on the control-flow (without conditions on data and time), and another one including real multi-perspective constraints (with conditions on time and data). We followed the same procedure to create models with 20, 30, 40, and 50 constraints.
%
\begin{table}
	\centering
	\begin{small}
	\begin{tabular}{rr|rrrr}
		\toprule
			&&	\multicolumn{4}{c}{\textbf{Number of log traces}} \\
			&&	\textbf{25\,000}	&	\textbf{50\,000}	&	\textbf{75\,000}	&	\textbf{100\,000} \\
		\midrule
		\multirow{5}{*}{\rotatebox[origin=c]{90}{\parbox{1.5cm}{\textbf{Events per trace}}}}
		& \textbf{10}	&	250\,000	&	500\,000	&	750\,000	&	1\,000\,000 \\
		& \textbf{20}	&	500\,000	&	100\,0000	&	1\,500\,000	&	2\,000\,000 \\
		& \textbf{30}	&	750\,000	&	150\,0000	&	2\,250\,000	&	3\,000\,000 \\
		& \textbf{40}	&	100\,0000	&	200\,0000	&	3\,000\,000	&	4\,000\,000 \\
		& \textbf{50}	&	125\,0000	&	250\,0000	&	3\,750\,000	&	5\,000\,000 \\
		\bottomrule
	\end{tabular}
	\end{small}
	\caption{Number of events for each log.}
	\label{tbl:events-per-log}
\end{table}

We checked each log against each model, and we repeated the procedure five times, in order to get the average execution times for each configuration. To provide more accurate results, the times reported here are measured without considering the time needed to generate the graphical visualization (we perform the tests on a custom command-line version of ProM).
All tests have been performed using two machines (part of a cluster) randomly, with the following hardware configurations:
	\emph{(i)} 4 x Eight-Core Intel(R) Xeon(R) CPU E5-4640 0 @ 2.40GHz;
	\emph{(ii)} 2 x Intel(R) Xeon(R) CPU E5-2670 0 @ 2.60GHz.

\begin{figure}[t!]
	\centering
	\includegraphics[width=.7\textwidth]{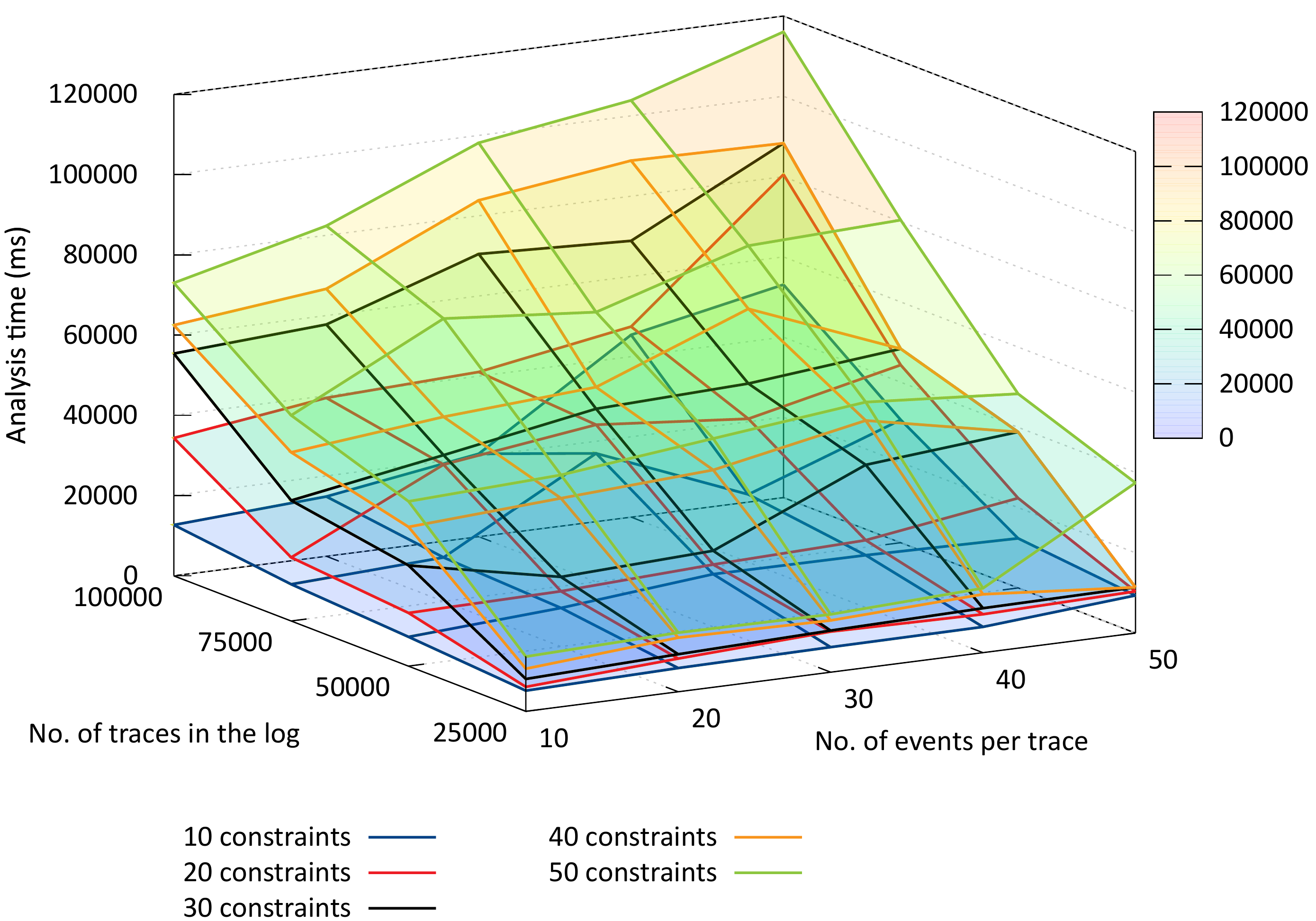}
	\includegraphics[width=.7\textwidth]{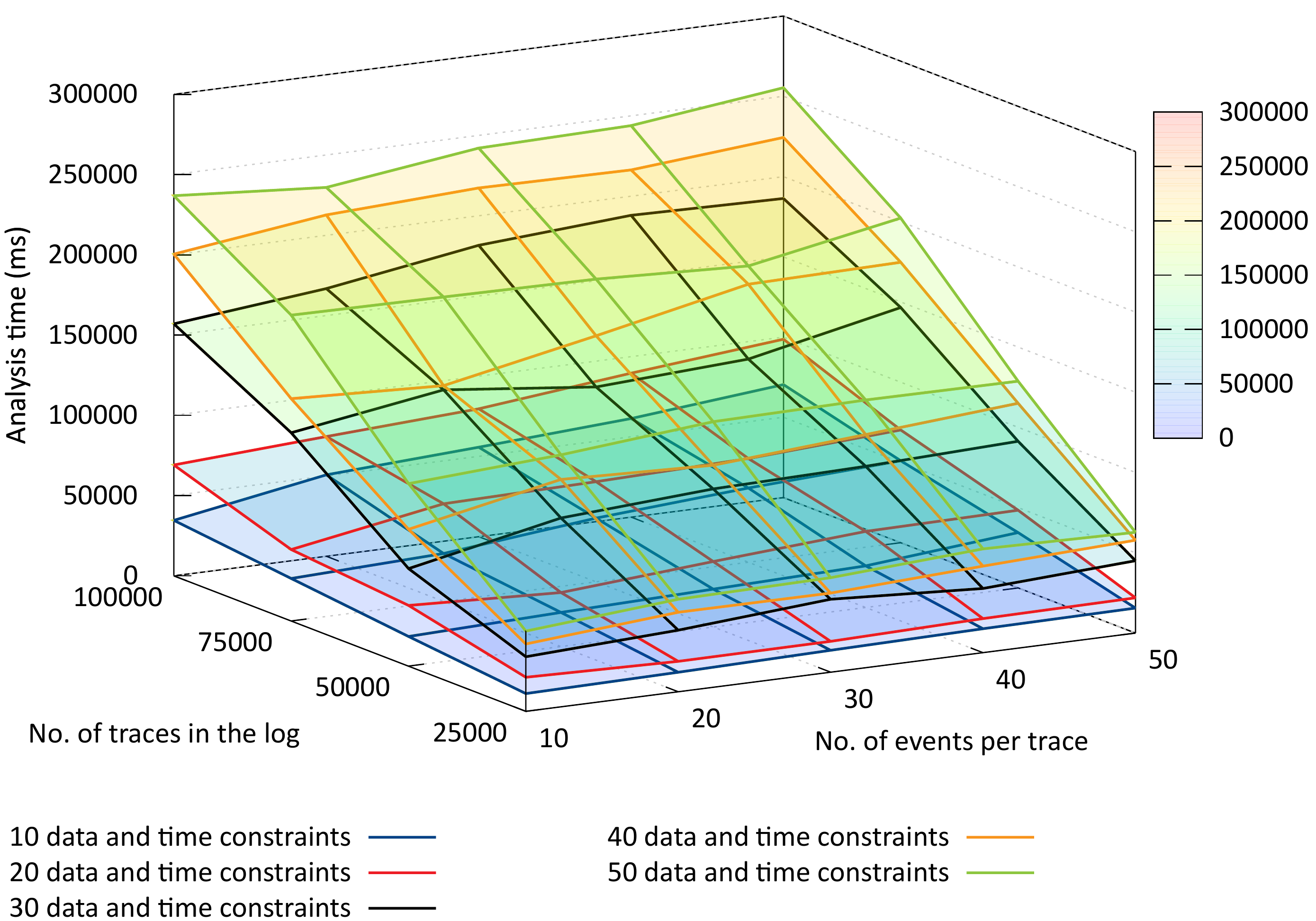}
	\caption{Execution times in milliseconds required to process logs with different number of traces of different lengths. The plot on top refers to models with control-flow constraints. The plot at the bottom refers to models with control-flow, data and time constraints.}
	\label{fig:benchmark}
\end{figure}

\begin{figure}[t!]
	\centering
	\includegraphics[width=.45\textwidth]{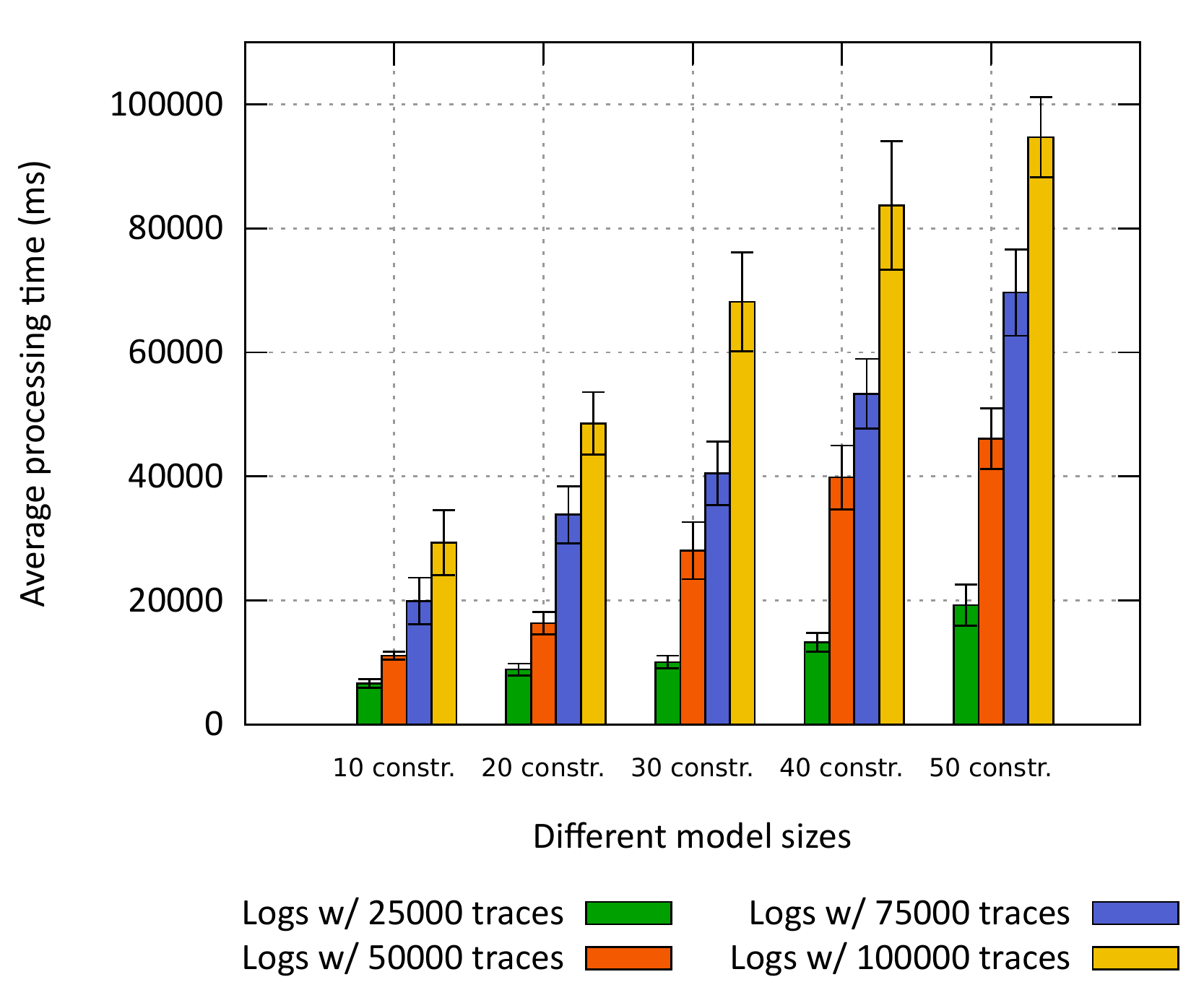}%
	\includegraphics[width=.45\textwidth]{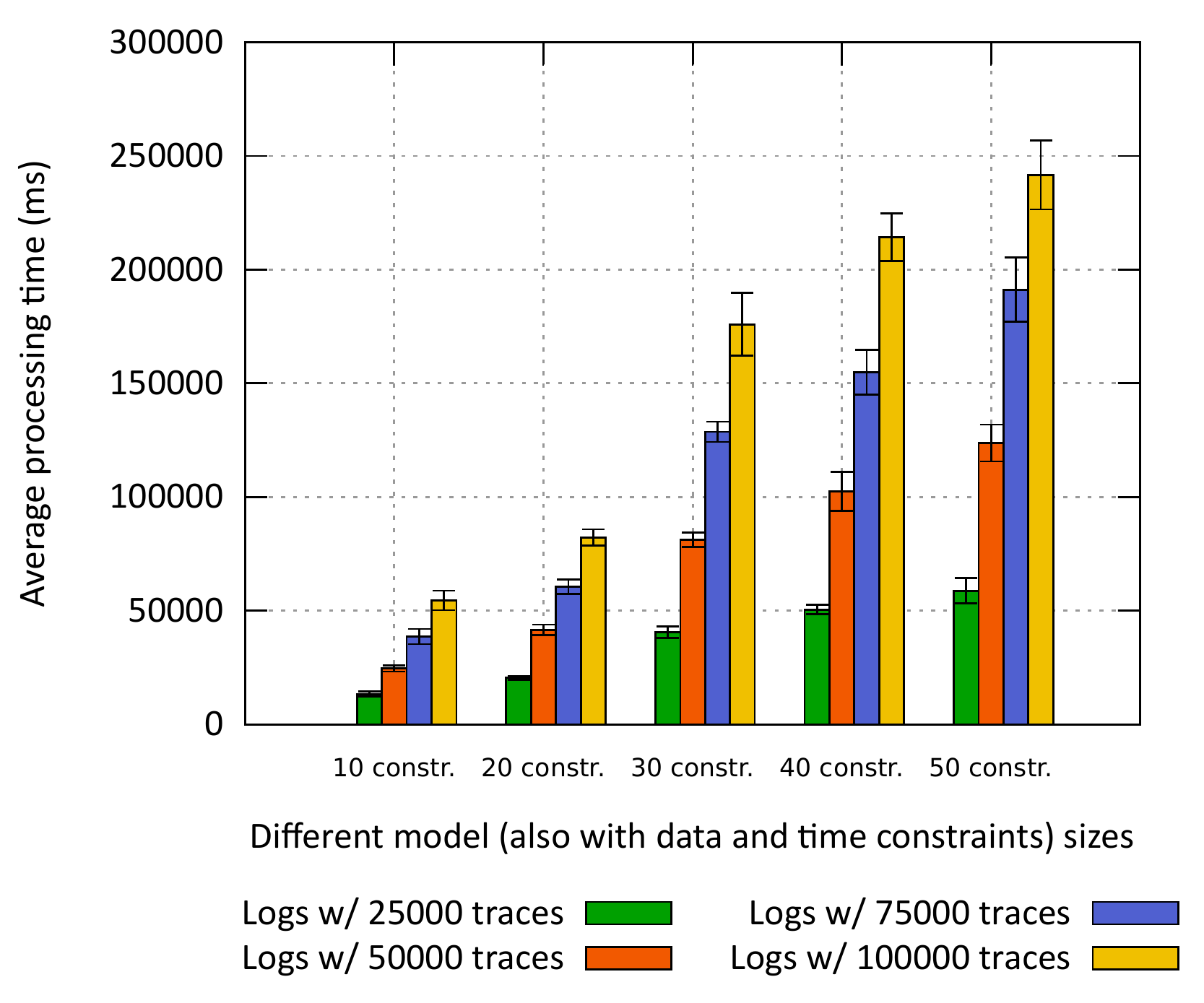}
	\caption{Execution times in milliseconds grouped based on the number of traces in the logs. The plot on the left hand side refers to models with control-flow constraints. The plot on the right hand side refers to models with control-flow, data and time constraints. }
	\label{fig:benchmark:averages1}
\end{figure}

\begin{figure}[t!]
	\centering
	\includegraphics[width=.45\textwidth]{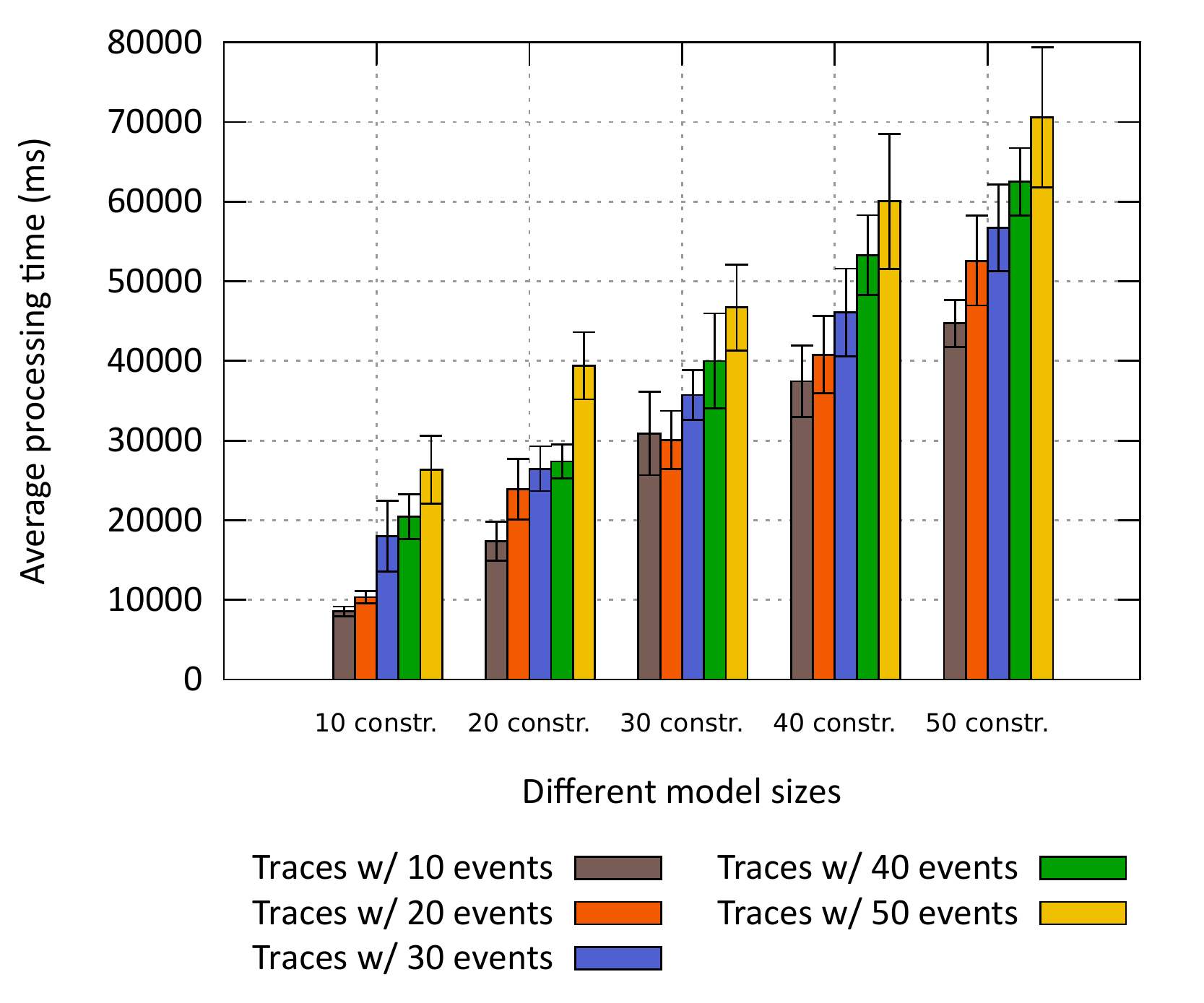}%
	\includegraphics[width=.45\textwidth]{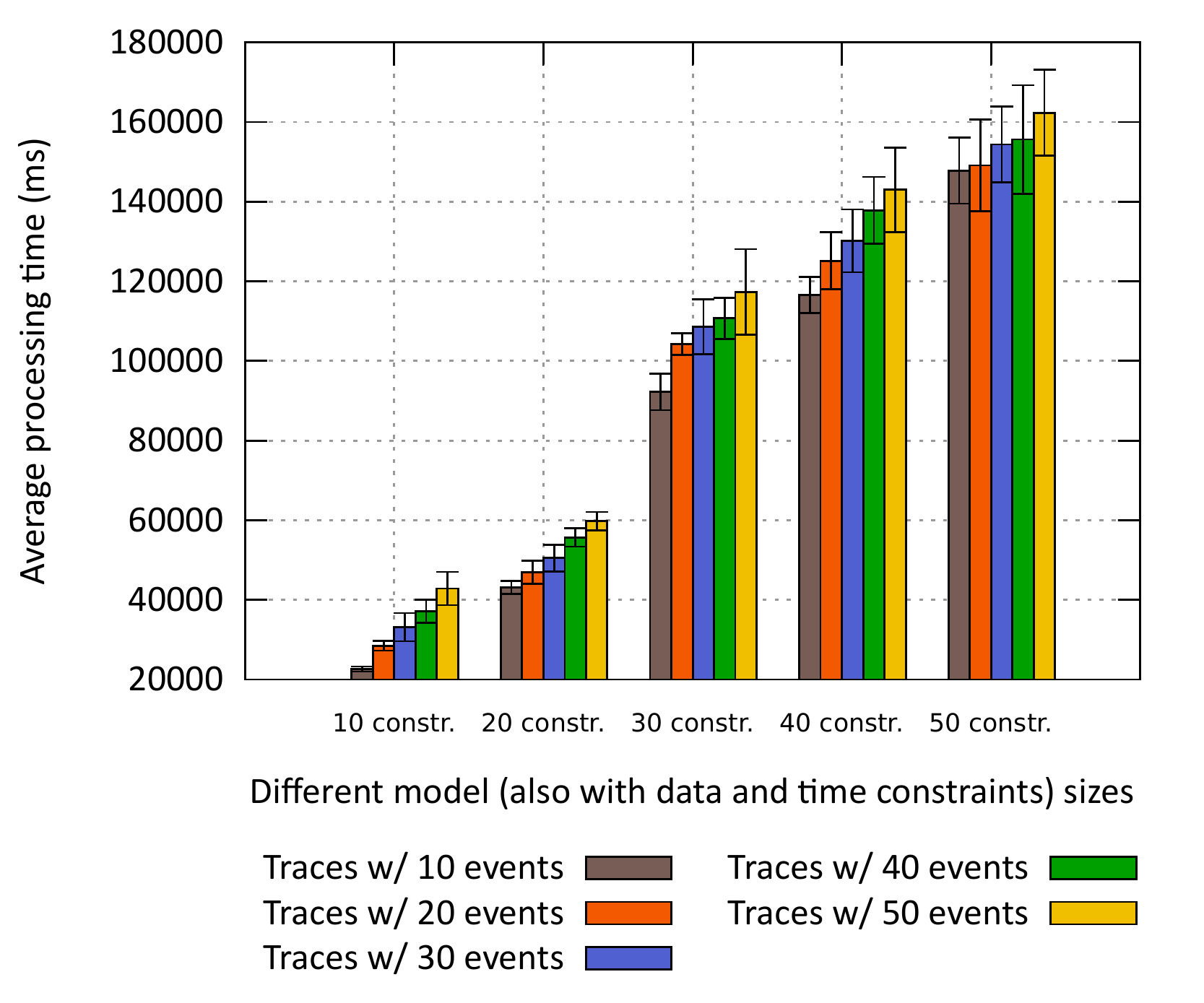}
	\caption{Execution times in milliseconds grouped based on the number of events in each trace. The plot on the left hand side refers to models with control-flow constraints. The plot on the right hand side refers to models with control-flow, data and time constraints. }
	\label{fig:benchmark:averages2}
\end{figure}

\figurename~\ref{fig:benchmark} provides a graphical representation of the average execution times for the analysis of all models and logs. In particular, the graph on top reports the execution times using models with control-flow based constraints. The graph at the bottom reports the execution times using real multi-perspective models (with conditions on time and data). In \figurename~\ref{fig:benchmark:averages1} and in \figurename~\ref{fig:benchmark:averages2}, we also report the average execution times (and standard deviations) required to analyze all models and logs but we provide different views on the data. In particular, in \figurename~\ref{fig:benchmark:averages1}, the execution times are grouped based on the number of traces in the logs. The graph on the left-hand side reports the execution times using models with control-flow based constraints, the one on the right-hand side reports the execution times using multi-perspective constraints. In \figurename~\ref{fig:benchmark:averages2}, the execution times are
grouped based on the number of events in each trace.

As the statistics clearly show, the time required to perform the analysis directly depends both on the number of events in each trace, and on the actual size of the log.
However, the execution times evaluated using models with control-flow based constraints seem to be more influenced by the number of events in each traces.
We believe that this is due to the additional costs needed for starting up the data validation engine in case of multi-perspective models. In particular, it is necessary to restart such engine for each trace and the additional time required is so high that it becomes impossible to see the differences in terms of performances for traces of different lengths.
In general, it is worthwhile noting that the most expensive configuration (a model with 50 multi-perspective constraints, and a log with 100\,000 traces and 5\,000\,000 events) requires, on average, 255\,369 milliseconds, i.e., about 4.2 minutes. This proves the scalability of our approach.

%% file: sections/casestudy.tex
\section{Case Studies}
\label{sec:casestudy}

This section provides three case studies on real datasets. The first one is based on an event log provided by an academic hospital, the second one is a case study provided by a financial institution and the third one is based on a dataset provided by a bank.

\input{sections/casestudy-hosp}

\input{sections/casestudy-financ}
\input{sections/casestudy-rabo}

%% file: sections/casestudy-hosp.tex
\begin{table}[t!]
\caption{Reference constraints used to analyze the log from the BPI challenge 2011.}
\centering
\scalebox{0.65}{
\begin{tabular}{l|l|p{2.5cm}p{2.5cm}|p{3cm}p{3cm}|l}
	\toprule
	\textbf{Id} &
		\textbf{Constraint} &
		\textbf{1st param.} &
		\textbf{2nd param.} &
		\parbox{3cm}{\textbf{Activation \\ condition}} &
		\parbox{3cm}{\textbf{Correlation \\ condition}} &
		\parbox{1.5cm}{\textbf{Time \\ condition}} \\[.5em]
	\midrule
	1 & Precedence &
		\texttt{ca-125 using meia} &
		\texttt{outpatient follow-up consultation} &
		\texttt{A.Diagnosis == 'maligniteit ovarium or tuba'} &
		\texttt{-} &
		\texttt{0,15,d} \\
	\midrule
	2 & Precedence &
		\texttt{First outpatient consultation} &
		\texttt{telephone consultation} &
		\texttt{-} &
		\texttt{A.org:group == T.org:group} &
		\texttt{-} \\
	\bottomrule
\end{tabular}
}
\label{tab:constr2011}
\end{table}
\begin{table}[t!]
\caption{Conformance checking results using the log from the BPI challenge 2011.}
\centering
\scalebox{0.65}{
\begin{tabular}{c|rrr|rrr}
	\toprule
	\textbf{Id} & \textbf{Act.no.} & \textbf{Viol.no.} & \textbf{Fulfill.no.} & \textbf{Avg.act.sparsity} & \textbf{Avg.viol.ratio} & \textbf{Avg.fulfill.ratio} \\
	\midrule
	1 & 343 & 242 & 101 & 0.9844 & 0.7055 & 0.2945 \\
	2 & 1\,286 & 546 & 740 & 0.9677 & 0.4246 & 0.5754 \\
	\bottomrule
\end{tabular}
}
\label{tab:confconstr2011}
\end{table}
\begin{table}[t!]
\caption{Execution times using the log from the BPI challenge 2011.}
\centering
\scalebox{0.65}{
\begin{tabular}{c|r}
	\toprule
	\textbf{Id} & \textbf{Avg.execution time (milliseconds)} \\
	\midrule
	1 & 1\,759  \\
	2 &  1\,828 \\
	\bottomrule
\end{tabular}
}
\label{tab:times2011}
\end{table}
\subsection{A Large Academic Hospital}
We have conducted a case study by using the BPI challenge 2011 event log \cite{bpichallenge2011}. This log pertains to a healthcare process and, in particular,
contains the executions of a process related to the treatment of patients diagnosed with cancer in a large Dutch academic hospital. The whole event log contains $1\,143$ cases and $150\,291$ events distributed across $623$ event classes (i.e., each event refers to one of $623$ different possible activities). Each case describes the treatment of a different patient. The event log contains domain specific attributes that are both case attributes and event attributes in addition to the standard XES attributes. For example, \texttt{Age}, \texttt{Diagnosis}, and \texttt{Treatment code} are case attributes and \texttt{Activity code}, \texttt{Number of executions}, \texttt{Specialism code}, and \texttt{Group} are event attributes. As mentioned in Section~\ref{sec:mining}, in our analysis all the attributes are considered visible for all the activities and we suppose that an activity overwrites the old values of all the event attributes attached to it.

\begin{figure}[t!]
    \centering
    \includegraphics[scale=0.5]{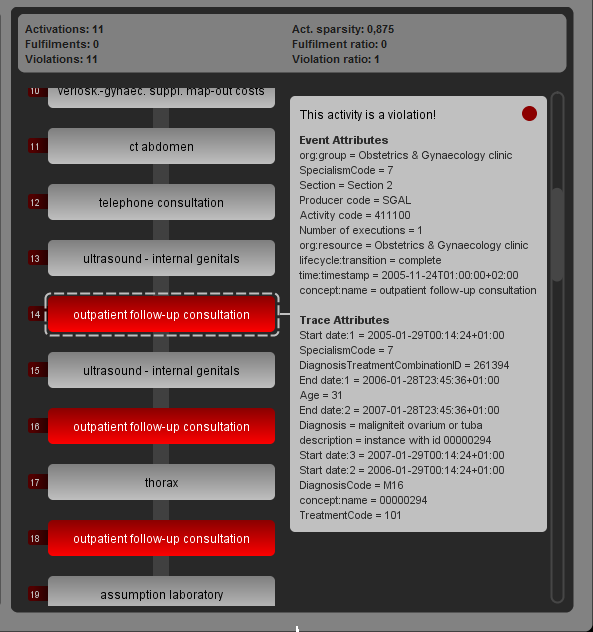}
    \caption{Example of violations for constraint 1.}\label{fig:prec2011Viol2}
\end{figure}

To investigate the behavior of the process as recorded in the log, we have used the constraints shown in \tablename~\ref{tab:constr2011}. The idea behind constraint 1 is that the tumor marker ``ca-125'' is used in the follow-up of patients diagnosed with ovarian cancer as an indicator of the evolution of the tumor. For this reason, we would expect that, if the diagnosis for a patient is ``maligniteit ovarium'', the follow-up consultation is preceded by the analysis of this tumor marker. In addition, we require a time condition indicating that this analysis should not come too early with respect to the follow-up. As shown in \tablename~\ref{tab:confconstr2011}, constraint 1 has 343 activations.
This means that there are 343 occurrences of \texttt{outpatient follow-up consultation} associated with a \texttt{Diagnosis} equal to \texttt{maligniteit ovarium or tuba}. As shown in \tablename~\ref{tab:confconstr2011}, around $70\%$ of these activations are violations. One of the reasons why there are so many violations in the log for this constraint is that there can be several follow-ups in a case and some of them are not correlated with the ``ca-125'' test but with other tests. In \figurename~\ref{fig:prec2011Viol2}, it is possible to see some violations for constraint 1. For example, the selected event \texttt{outpatient follow-up consultation} is an activation for the constraint since, in its payload, the value for \texttt{Diagnosis} is \texttt{maligniteit ovarium or tuba}. However, this activation is probably connected with the computed tomography abdomen and/or the ultrasound test done immediately before.

The idea behind constraint 2 is that the first consultation for a patient in the hospital cannot be a telephone consultation. We also add a correlation condition to understand if every telephone consultation is preceded by a first consultation in the same department. There is no activation condition for this constraint. This means that every time \texttt{telephone consultation} occurs, the constraint is activated. The constraint has 1\,286 activations. Around $42\%$ of these activations are violations. Some of these violations are due to the occurrence of telephone consultations preceded by a first consultation in a different department. In addition, it is also worth to highlight that the log we are using for this case study is an excerpt derived from a larger log and it contains several cases that are truncated both at the beginning and at the end. This can be also the reason of violations for this constraint.

In Table~\ref{tab:times2011}, we show the execution times needed for checking the constraints in this case study.\footnote{The execution times in all the tables of this section are averaged over 5 runs.} \footnote{All the experiments described in this section have been performed on a machine with an Intel(R) Core(TM) i7-2670QM CPU @ 2.20GHz (limiting the execution to just one core), 8 GB of RAM and the Oracle Java virtual machine installed on a GNU/Linux Ubuntu operating system.} For each of them, the execution time is lower that 2 seconds. This confirms that the scalability of our tool.

%% file: sections/casestudy-financ.tex
\begin{table}[b!]
\caption{Reference constraints used to analyze the log from the BPI challenge 2012.}
\centering
\scalebox{0.7}{
\begin{tabular}{l|l|p{3cm}p{3cm}|p{2cm}p{2.5cm}|l}
	\toprule
	\textbf{Id} &
		\textbf{Constraint} &
		\textbf{1st param.} &
		\textbf{2nd param.} &
		\parbox{3cm}{\textbf{Activation \\ condition}} &
		\parbox{3cm}{\textbf{Correlation \\ condition}} &
		\parbox{1.5cm}{\textbf{Time \\ condition}} \\[.5em]
	\midrule
		3 & Response &
		\texttt{A\_SUBMITTED} &
		\texttt{A\_ACCEPTED} &
		\texttt{-} &
		\texttt{-} &
		\texttt{-} \\
	\midrule
		4 & Response &
		\texttt{A\_SUBMITTED} &
		\texttt{A\_ACCEPTED} &
		\texttt{-} &
		\texttt{-} &
		\texttt{0,24,h} \\
	\midrule
		5 & Response &
		\texttt{A\_SUBMITTED} &
		\texttt{A\_ACCEPTED} &
     	\texttt{A.AMOUNT\_REQ >= 10\,000} &
		\texttt{-} &
		\texttt{-} \\
\midrule
		6 & Response &
		\texttt{A\_SUBMITTED} &
		\texttt{A\_ACCEPTED} &
     	\texttt{A.AMOUNT\_REQ < 10\,000} &
		\texttt{-} &
		\texttt{-} \\	
\midrule
		7 & Response &
		\texttt{W\_Valideren aanvraag-SCHEDULE} &
		\texttt{W\_Valideren aanvraag-START} &
		\texttt{-} &
		\texttt{-} &
		\texttt{-} \\
\midrule
		8 & Response &
		\texttt{W\_Valideren aanvraag-SCHEDULE} &
		\texttt{W\_Valideren aanvraag-START} &
		\texttt{-} &
		\texttt{A.org:resource != T.org:resource} &
		\texttt{-} \\
\midrule
		9 & Response &
		\texttt{W\_Valideren aanvraag-SCHEDULE} &
		\texttt{W\_Valideren aanvraag-START} &
		\texttt{-} &
		\texttt{A.org:resource != T.org:resource} &
		\texttt{0,7,d} \\
\midrule
		10 & Response &
		\texttt{W\_Valideren aanvraag-SCHEDULE} &
		\texttt{W\_Valideren aanvraag-START} &
		\texttt{-} &
		\texttt{A.org:resource != T.org:resource} &
		\texttt{0,24,h} \\
\midrule
		11 & Response &
		\texttt{W\_Valideren aanvraag-START} &
		\texttt{W\_Valideren aanvraag-COMPLETE} &
		\texttt{-} &
		\texttt{-} &
		\texttt{-} \\
\midrule
		12 & Response &
		\texttt{W\_Valideren aanvraag-START} &
		\texttt{W\_Valideren aanvraag-COMPLETE} &
		\texttt{-} &
		\texttt{A.org:resource == T.org:resource} &
		\texttt{-} \\
\midrule
		13 & Response &
		\texttt{W\_Valideren aanvraag-START} &
		\texttt{W\_Valideren aanvraag-COMPLETE} &
		\texttt{-} &
		\texttt{A.org:resource == T.org:resource} &
		\texttt{0,1,h} \\
\midrule
		14 & Response &
		\texttt{W\_Valideren aanvraag-START} &
		\texttt{W\_Valideren aanvraag-COMPLETE} &
		\texttt{-} &
		\texttt{A.org:resource == T.org:resource} &
		\texttt{0,15,m} \\
	\bottomrule
  \end{tabular}
}
\label{tab:constr2012}
\end{table}
\begin{table}[b!]
\caption{Conformance checking results using the log from the BPI challenge 2012.}
\centering
\scalebox{0.65}{
\begin{tabular}{c|rrr|rrr}
	\toprule
	\textbf{Id} & \textbf{Act.no.} & \textbf{Viol.no.} & \textbf{Fulfill.no.} & \textbf{Avg.act.sparsity} & \textbf{Avg.viol.ratio} & \textbf{Avg.fulfill.ratio} \\
	\midrule
	3 & 13\,087 & 7\,974 &   5\,113 & 0.8596 & 0.6093 & 0.3907 \\
	4 & 13\,087 &  9\,036 & 4\,051 & 0.8596 & 0.6905 & 0.3095 \\
	5 &  6\,847 & 3\,601 &  3\,246 & 0.9585 & 0.5259 & 0.4741 \\
	6 &  6\,240 & 4\,373 &   1\,867 & 0.9211 & 0.7008 & 0.2992 \\
	7 &  5\,023 & 51 &   4\,972 & 0.9909 & 0.0102 & 0.9898 \\
	8 & 5\,023 &  236 & 4\,787 & 0.9909 & 0.047 & 0.953 \\
	9 &  5\,023 & 263 &  4\,760 & 0.9909 & 0.0524 & 0.9476 \\
	10 &  5\,023 & 2\,897 &   2\,126 & 0.9909 & 0.5767 & 0.4233 \\
	11 &  7\,891 & 2 &   7\,889 & 0.9863 & 0.0003 & 0.9997 \\
    12 &  7\,891 & 6 &   7\,885 & 0.9863 & 0.0008 & 0.9992 \\	
    13 & 7\,891 &  228 & 7\,663 & 0.9863 & 0.0289 & 0.9711 \\
	14 &  7\,891 & 3\,355 &   4\,536 & 0.9863 & 0.4252 & 0.5748 \\
	\bottomrule
\end{tabular}
}
\label{tab:confconstr2012}
\end{table}
\begin{table}[b!]
\caption{Execution times using the log from the BPI challenge 2012.}
\centering
\scalebox{0.65}{
\begin{tabular}{c|r}
	\toprule
	\textbf{Id} & \textbf{Avg.execution time (milliseconds)} \\
	\midrule
	3 & 2\,772 \\
	4 & 3\,220 \\
	5 &  3\,261  \\
	6 &  3\,205  \\
	7 &  3\,196  \\
	8 & 3\,100  \\
	9 &  3\,212  \\
	10 &  3\,146  \\
	11 &  2\,176  \\
    12 &  3\,210  \\	
    13 & 3\,241  \\
	14 &  3\,258  \\
	\bottomrule
\end{tabular}
}
\label{tab:exetime2012}
\end{table}





\begin{figure}[b!]
	\centering
	\begin{subfigure}[b]{0.49\textwidth}
		\includegraphics[width=\textwidth]{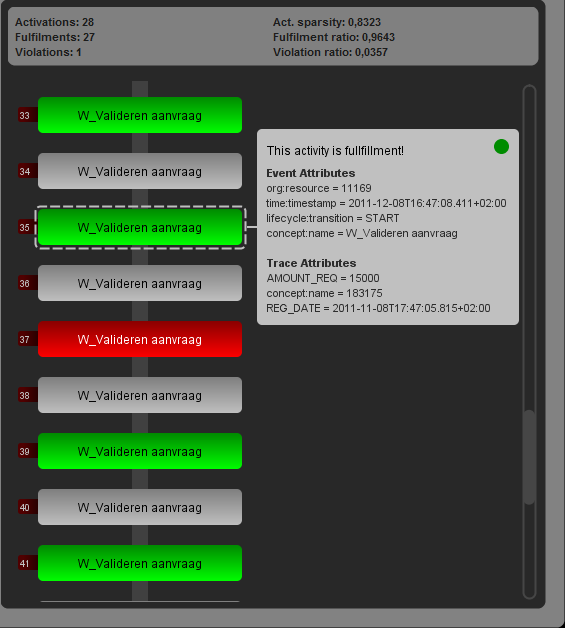}
		\caption{Example of fulfillment \texttt{W\_Valideren aanvraag-START} at position 35. \\~}
		\label{fig:prec2012Viol1}
	\end{subfigure}
	\begin{subfigure}[b]{0.49\textwidth}
		\includegraphics[width=\textwidth]{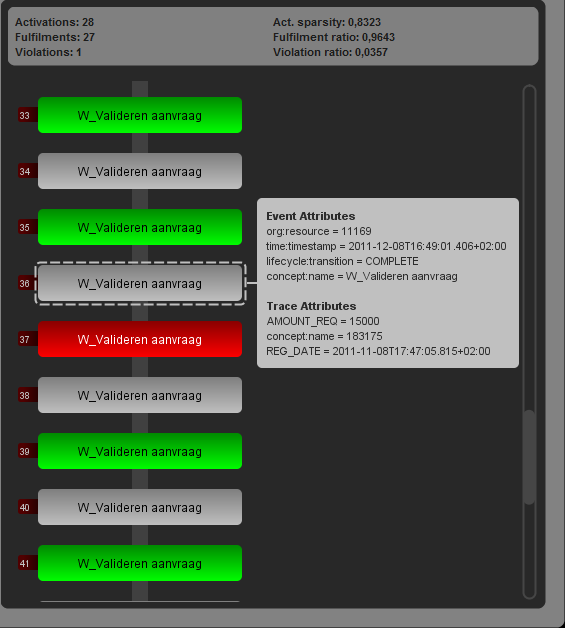}
		\caption{A correlated target \texttt{W\_Valideren aanvraag-COMPLETE} at position 36 executed by the same resource.}
		\label{fig:prec2012Viol2}
	\end{subfigure}
	\caption{Example of fulfillment for constraint 13.}
	\label{fig:2012_1}
\end{figure}

\begin{figure}[t!]
	\centering
	\begin{subfigure}[b]{0.49\textwidth}
		\includegraphics[width=\textwidth]{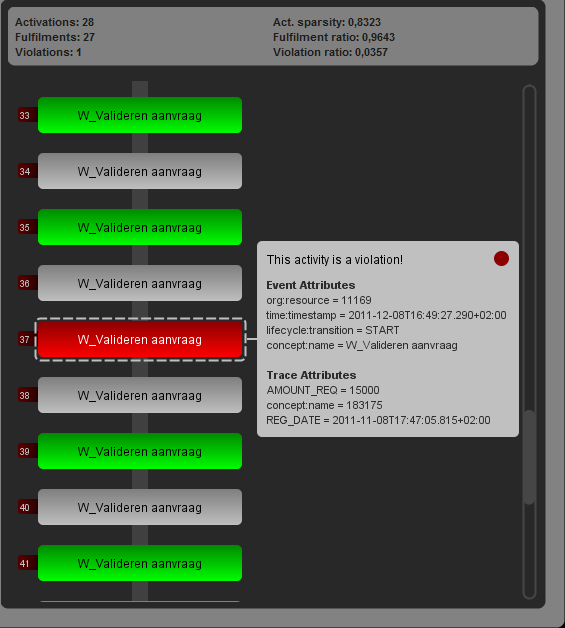}
		\caption{Example of violation \texttt{W\_Valideren aanvraag-START} at position 37.\\~}
		\label{fig:prec2012Viola1}
	\end{subfigure}
	\begin{subfigure}[b]{0.49\textwidth}
		\includegraphics[width=\textwidth]{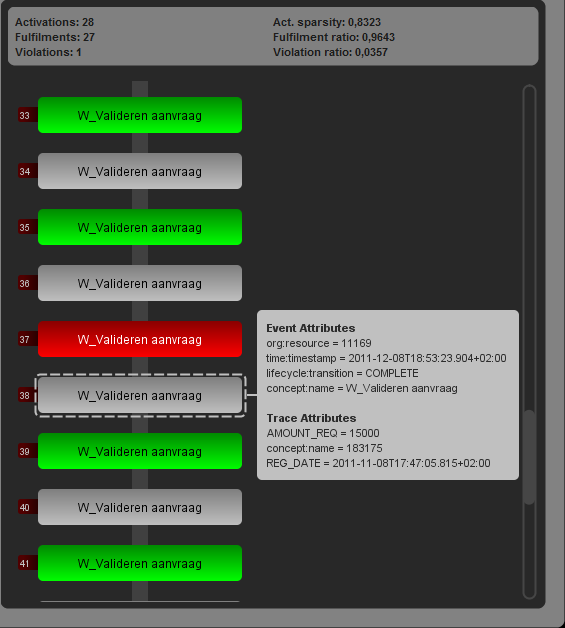}
		\caption{A possible target \texttt{W\_Valideren aanvraag-COMPLETE} occurs more than 1 hours after.}
		\label{fig:prec2012Viola2}
	\end{subfigure}
	\caption{Example of violation for constraint 13; \texttt{W\_Valideren aanvraag-COMPLETE} occurs outside the required time interval (too late).}
	\label{fig:2012_2}
\end{figure}

\begin{figure}[t!]
	\centering
	\begin{subfigure}[b]{0.49\textwidth}
		\includegraphics[width=\textwidth]{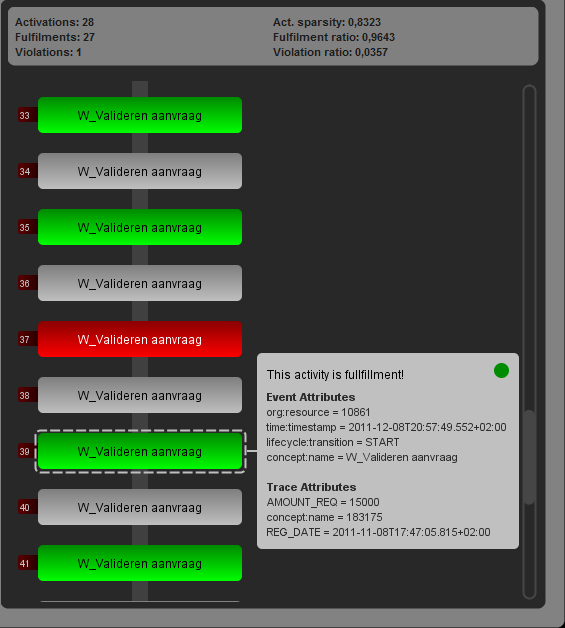}
		\caption{Example of fulfillment \texttt{W\_Valideren aanvraag-START} at position 39.}
		\label{fig:prec2012Fulfill1}
	\end{subfigure}
	\begin{subfigure}[b]{0.49\textwidth}
		\includegraphics[width=\textwidth]{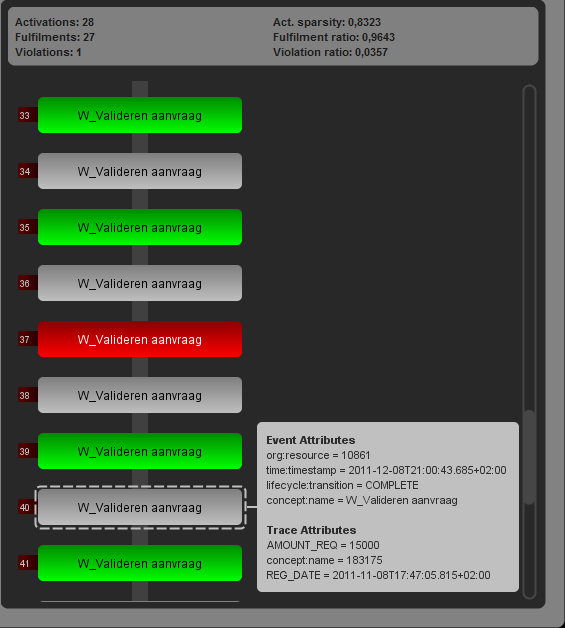}
		\caption{Corresponding target executed by the same resource.}
		\label{fig:prec2012Fulfill2}
	\end{subfigure}
	\caption{Example of fulfillment for constraint 13; \texttt{W\_Valideren aanvraag-START} at position 39 is followed by \texttt{W\_Valideren aanvraag-COMPLETE} within the required time interval.}
	\label{fig:2012_3}
\end{figure}

\subsection{A Dutch Financial Institution}
The second case study we discuss is based on the application of the proposed approach to the event log provided for the BPI challenge 2012 and taken from a Dutch financial institute \cite{bpichallenge2012}. The event log pertains to an application process for personal loans or overdrafts. It contains 262\,200 events distributed across 36 event classes and includes 13\,087 cases. The amount requested by the customer is indicated in the case attribute \texttt{AMOUNT\_REQ}. In addition, the log contains the standard XES attributes for events. 

For this case study, we have used the constraints shown in \tablename~\ref{tab:constr2012}. Some of these constraints involve some specific transactional states (a.k.a. event types) of an activity. For example, the parameters specified for constraint 7-10 are \texttt{W\_Valideren aanvraag-SCHEDULE} and \texttt{W\_Valideren aanvraag-START}. When an event type is not specified, like in the case of constraint 3-6, the event type considered by default is ``complete''.

With constraint 3, we want to understand how many submitted applications are eventually accepted. As shown in \tablename~\ref{tab:confconstr2012}, there are 13\,087 submissions of which only 5\,113 are eventually accepted (around $39\%$). Using constraint 4, we can understand that the majority of these accepted applications (around $79\%$) are accepted in less than 24 hours from the submission. Using constraints 5 and 6, we can understand how the requested amount affects the application. In particular, when the requested amount is lower than 10\,000 the acceptance rate is almost $30\%$. The acceptance rate is higher if the requested amount is greater or equal to 10\,000 (almost half of the applications is accepted in this case).

With constraints 7-14, we analyze the validation of the applications.
With constraint 7, we can see that almost $99\%$ of the scheduled validations are eventually started. In $95\%$ of the cases, the resource that schedules the validation is not the same resource that starts this activity (see constraint 8). In addition, in around $94\%$ of the cases, a scheduled validation is started within 7 days from the scheduling (constraint 9) and in almost half of the cases the validation is started only 24 hours after the scheduling. Constraint 11 indicates that almost $100\%$ of the validations that have been started are also completed, and almost in all the cases the resource that starts the validation is the same resource that completes this activity (see constraint 12). In $97\%$ of the cases, the validation is done in at most 1 hour (constraint 13), and in more than half of the cases it is completed in less than 15 minutes (constraint 14).

In \figurename~\ref{fig:2012_1} and \ref{fig:2012_3}, we show two fulfillments for constraint 13 (the activations with the correlated targets). \ref{fig:2012_3} shows a violation for the same constraint.
In Table~\ref{tab:exetime2012}, we show the execution times needed for checking the constraints in this case study. Also in this case, like in the first case study here presented, the execution time is low (between 2 and 3 seconds on average). 

%% file: sections/casestudy-rabo.tex
\begin{table}[t!]
\caption{Reference constraints used to analyze the log from the BPI challenge 2014.}
\centering
\scalebox{0.65}{
\begin{tabular}{l|l|ll|p{3cm}p{3cm}|l}
	\toprule
	\textbf{Id} &
		\textbf{Constraint} &
		\textbf{1st param.} &
		\textbf{2nd param.} &
		\parbox{3cm}{\textbf{Activation \\ condition}} &
		\parbox{3cm}{\textbf{Correlation \\ condition}} &
		\parbox{1.5cm}{\textbf{Time \\ condition}} \\[.5em]
	\midrule
		15 & Not response &
		\texttt{Open} &
		\texttt{Reopen} &
		\texttt{-} &
		\texttt{-} &
		\texttt{-} \\
	\midrule
		16 & Not response &
		\texttt{Open} &
		\texttt{Reopen} &
		\texttt{-} &
		\texttt{A.org:resource != T.org:resource} &
		\texttt{-} \\
	\midrule
		17 & Response &
		\texttt{Open} &
		\texttt{Closed} &
		\texttt{-} &
		\texttt{-} &
		\texttt{-} \\
    \midrule
		18 & Response &
		\texttt{Open} &
		\texttt{Closed} &
		\texttt{-} &
		\texttt{-} &
		\texttt{0,12,h} \\
    \midrule
		19 & Response &
		\texttt{Open} &
		\texttt{Closed} &
		\texttt{A.KMnumber == 'KM0000611'} &
		\texttt{-} &
		\texttt{0,12,h} \\
	\midrule
		20 & Response &
		\texttt{Open} &
		\texttt{Closed} &
		\texttt{A.KMnumber == 'KM0002043'} &
		\texttt{-} &
		\texttt{0,12,h} \\
		
\bottomrule
  \end{tabular}
}
\label{tab:constr2014}
\end{table}
\begin{table}[t!]
\caption{Conformance checking results using the log from the BPI challenge 2014.}
\centering
\scalebox{0.65}{
\begin{tabular}{c|rrr|rrr}
	\toprule
	\textbf{Id} & \textbf{Act.no.} & \textbf{Viol.no.} & \textbf{Fulfill.no.} & \textbf{Avg.act.sparsity} & \textbf{Avg.viol.ratio} & \textbf{Avg.fulfill.ratio} \\
	\midrule
	15 & 46\,607 &  2\,121 & 44\,486 & 0.8468 & 0.0455 & 0.9545 \\
	16 & 46\,607 &   510 & 46\,097 & 0.8468 & 0.0109 & 0.9891 \\
	17 & 46\,607 &   449 & 46\,158 & 0.8468 & 0.0096 & 0.9904 \\
	18 & 46\,607 & 24\,392 & 22\,215 & 0.8468 & 0.5234 & 0.4766 \\
	19 &   446 &   386 &    60 & 0.9993 & 0.8655 & 0.1345 \\
	20 &   773 &    48 &   725 & 0.9969 & 0.0621 & 0.9379 \\	
\bottomrule
\end{tabular}
}
\label{tab:confconstr2014}
\end{table}

\begin{table}[b!]
\caption{Execution times using the log from the BPI challenge 2014.}
\centering
\scalebox{0.65}{
\begin{tabular}{c|r}
	\toprule
	\textbf{Id} & \textbf{Avg.execution time (milliseconds)} \\
	\midrule
	15 &  4\,294  \\
	16 &  5\,093  \\
	17 &  5\,240  \\
	18 &  5\,055  \\
	19 &  4\,861  \\
	20 &  5\,398  \\
	\bottomrule
\end{tabular}
}
\label{tab:exetime2014}
\end{table}

\begin{figure}[t!]
	\centering
	\begin{subfigure}[b]{0.49\textwidth}
		\includegraphics[width=\textwidth]{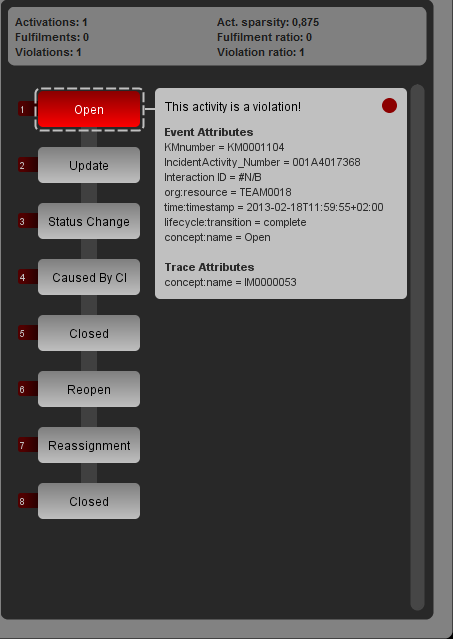}
		\caption{Example of violation \texttt{Open} at position 1.\\~}
		\label{fig:OpReoPResourceDiffViol1}
	\end{subfigure}
	\begin{subfigure}[b]{0.49\textwidth}
		\includegraphics[width=\textwidth]{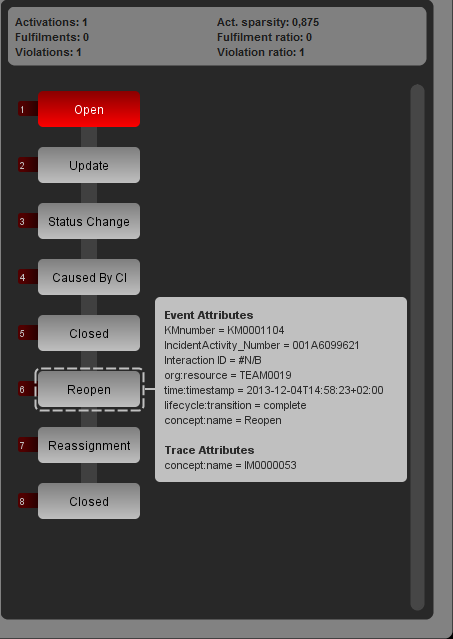}
		\caption{A forbidden event \texttt{Reopen} occurs after \texttt{Open}.}
		\label{fig:OpReoPResourceDiffViol2}
	\end{subfigure}
	\caption{Example of violation for constraint 16; \texttt{Open} is followed by an event \texttt{Reopen} associated to a different resource.}
	\label{fig:2014}
\end{figure}

\subsection{Rabobank}
The case study we illustrate in this section has been provided for the BPI challenge 2014 by Rabobank Netherlands Group ICT \cite{b4f76f0c-58a0-445b-b1ec-28fc470dcd9d}. The log we use pertains to the management of calls or mails from customers to the Service Desk concerning disruptions of ICT-services. The log contains 46\,616 cases, 466\,737 events referring to 39 different event classes. There are 242 originators and domain specific event attributes like \texttt{KM number}, \texttt{Interaction ID} and \texttt{IncidentActivity\_Number}. For this case study, we have used the constraints shown in \tablename~\ref{tab:constr2014}.

As shown in \tablename~\ref{tab:confconstr2014}, constraint 15 has 46\,607 activations and 44\,486 fulfillments. This allows us to understand that in around $95\%$ of open calls are not reopened afterwards. This percentage is even higher if we require that an open call cannot be eventually reopened by the same resource (see constraint 16). Indeed, this is true in almost $99\%$ of the cases.

Around $99\%$ of the open calls are eventually closed (see constraint 17). Around half of them are closed within 12 hours (constraint 18). The ``KM number'' in this case study identifies the characteristics of a call to understand how urgent the corresponding problem is. The checks on rules 19 and 20 show that the calls corresponding to the number \texttt{KM0002043} are, in general, more urgent than the ones corresponding to the number \texttt{KM0000611}. Indeed, over $446$ calls corresponding to the KM number \texttt{KM0000611} only $60$ are closed within 12 hours. On the other hand, over $773$ calls corresponding to the KM number \texttt{KM0002043}, $725$ are closed within 12 hours.

\figurename~\ref{fig:2014} shows a violation for constraint 16. The selected event \texttt{Open} is followed by a forbidden event \texttt{Reopen} (associated to a different resource). Table~\ref{tab:exetime2014} shows that the execution times for this case study range from 4 to 5 seconds.

%% file: sections/conclusion.tex
\section{Conclusion and Future Work}
\label{sec:conclusion}

In this work, we propose a framework for checking the conformance of event logs with respect to \MPDeclare{} models. \MPDeclare{} is an extension of the declarative process modeling language Declare that allows the modeler to specify constraints over the data associated to the control-flow and over the ``time dimension'' of a business process. We describe and discuss in detail how the proposed framework can be used to define algorithms for conformance checking based on \MPDeclare{}.
Our proposal has been implemented in the process mining tool ProM. The implemented software covers the entire set of \MPDeclare{} templates. In addition, the conformance checker can also be used with standard Declare. A wide experimentation has been carried out using both real-life and synthetic logs. These case studies prove the applicability of our implementation in realistic settings.
Although it is extremely important to recognize deviances \textit{a-posteriori}, in some particular contexts, it would be also useful to detect violations on-the-fly as they occur. To this aim, in the near future we are planning to make the proposed framework suitable to be used in online settings.